\documentclass[11pt]{article}

\usepackage[parsep]{collref}

\usepackage{hyperref}
\hypersetup{
    colorlinks=true,
    linkcolor=blue,
    filecolor=magenta,      
    urlcolor=blue,
    citecolor=blue,
    linktocpage=true,
    }

\usepackage{latexsym,amsmath,amssymb,theorem,epsfig}
\DeclareFontFamily{OT1}{rsfs10}{}
\DeclareFontShape{OT1}{rsfs10}{m}{n}{ <-> rsfs10 }{}
\DeclareMathAlphabet{\mathscript}{OT1}{rsfs10}{m}{n}

\numberwithin{equation}{section}

\newcommand{\RR}{{\mathbf{R}}}

\newcommand{\com}[2]{[#1,#2]}

\def\a{\alpha}
\def\b{\beta}
\def\g{\gamma}

\def\r{\rho}
\def\s{\sigma}
\def\t{\tau}

\def\G{\Gamma}

\def\P{\Pi}

\def \S {{\cal S}}

\def\BB{\beta}
\def\GG{\gamma}

\def\gsim{ \lower .75ex \hbox{$\sim$} \llap{\raise .27ex \hbox{$>$}} }
\def\lsim{ \lower .75ex \hbox{$\sim$} \llap{\raise .27ex \hbox{$<$}} }
\def\be{\begin{equation}}
\def\ee{\end{equation}}
\def\bea{\begin{eqnarray}}
\def\eea{\end{eqnarray}}

\def \ha {\tfrac{1}{2}}

\def \del{\partial}
\def \a {\alpha}

\def\ov{\over}
\def \ci {\cite}

\def \foot {\footnote}

\def\la{\label}
\def\foot{\footnote}\newcommand{\rf}[1]{(\ref{#1})}
\def \OO {{\cal  O}}\def \no {\nonumber}

\def \adss {$AdS_5 \times S^5$\ }

\def \g {{\gamma} } \def \hD {\hat \D}
\def \ov {\over}
\def \ci {\cite}

\def \ed {
\bibliography{biblio.bib}
\bibliographystyle{JHEP-v2.9}
\end{document}
}

\def \edo {\end{document}}
\def \Tr {{\rm Tr}}

\usepackage{tikz}
\usetikzlibrary{decorations.markings}

\makeatletter
\renewcommand*{\@fnsymbol}[1]{\textit{\@alph{#1}}}
\makeatother

\begin{document}
%%%%%%%%%%%%%%%%%%%%%%%%%%%%%%%%%%%%%%%%%%%%%%%%%%%%%%%%%%%%%%%%%%%%%%
\begin{titlepage}
\vspace{-15cm}
%\today
%\hfill
%Imperial-TP-FS-2023-??
%\vskip-1pt
\vspace{-8cm}
\title{\vspace{-3cm}
  \hfill{\small Imperial-TP-FS-2023-01  }  \\ \vspace{1cm} %[3em]
   { %\LARGE 
   S-matrix  on  effective string % \\[0.05em]
    \\  
   and compactified  membrane}
   %   models with infinite tower of massive modes...
   % Adding mass to the effective string
  \\[1em] }

%\vspace{3cm}
\author{\large Fiona K. Seibold\thanks{f.seibold21@imperial.ac.uk} \, 
    and   Arkady A. Tseytlin\footnote{Also at  the Institute  for Theoretical and Mathematical Physics    (ITMP)  of  MSU
      and  Lebedev
    Institute.}  % \\  %%}\footnote{
   %\ \ \ \ \ \ \  \hspace {2 cm}   \ \ \ \ \  tseytlin@ic.ac.uk}
      \\[1.3em]
    {   Blackett Laboratory, Imperial College, London SW7 2AZ, U.K. }
}
%\date{}\today
  \maketitle
  %%%%%%%%%%%%%%%%%%%%%%%%%%%%
{\small
\begin{abstract}
{\small
Expanding   Nambu-Goto   action near infinitely long string vacuum  %and using the static gauge 
  one   can compute   scattering  amplitudes of  2d massless  fields   representing  transverse string coordinates. 
  As was  shown in  arXiv:1203.1054, the 
    resulting S-matrix is integrable  (provided appropriate local counterterms are added), 
  in agreement with  known free string  spectrum   and also with an interpretation of the static-gauge  NG   action as a $T\bar T$ deformation of a free  massless   theory.  
  We  consider a generalization of this computation  to the case of a membrane,  expanding its 3d action 
  near an infinite membrane vacuum that  has cylindrical $\mathbb R \times S^1$   shape (we refer to such membrane as ``compactified").  Representing 3d fields  as Fourier series in $S^1$   coordinate we get  an  effective  2d 
  model     in which the massless   string modes  are coupled to an infinite KK tower of  massive  2d modes. 
  We find that the resulting  2d S-matrix   is not integrable  already  at the tree level. We also compute 1-loop 
   scattering amplitude  of     massless string modes   with all   compactified membrane modes propagating in the loop. 
   The  result is UV finite  and is a non-trivial  function of  the kinematic variables. 
   In the  large momentum limit or when  the radius of $S^1$    is taken to infinity 
   we recover the 
   expression   for the 1-loop scattering  amplitude of the  uncompactified   $\mathbb R^2$  membrane. 
   We also   consider  a 2d model  which is the $T\bar T$ deformation to the   free theory  with the same   massless  plus  infinite massive  tower of modes. The  corresponding 2d  S-matrix is found, as expected,  to be integrable.
% send Simone
}
\end{abstract}
}

\

%v3
 \hfill{\small  contribution to the special issue of Journal of Physics A:\qquad \qquad\qquad \qquad \qquad \qquad  }
 
   \hfill{\qquad  \small \it ``Fields, Gravity, Strings and Beyond: In Memory of Stanley Deser''  \qquad \qquad \qquad \qquad \qquad }

\end{titlepage}

\def \iffa  {\iffalse}
\def \RR {{\mathbb R}}
\def \R {{\rm R}}
\def \s {\sigma} \def \t {\tau} 
\def \G {\Gamma} \def \four {{1\ov 4}}
\def \CP  {{\rm CP}}\def \gs {g_s}
\def \hD   {\hat{D}}
\def \JJ {{\rm J}}
\def \te {\textstyle}
\def \zz  {^{(0)}} 
\def \ve {\varepsilon}
\def \zo {^{(1)}}
 \def \rR  {{\rm R}} 
 \def \eps {\epsilon}
 \def \rS  {{\rm S}} \def \rT {{\rm T}}
\def \P   {{\rm P}}
 \def \bbeta {{\rm r}} 
\def \no {\nonumber}
\def \TTT  {{\cal T}} 
\def \S {{\rm S}}  \def \rT  {{\cal T}} 
\def \dDelta {{\hat \Delta}}
\def \ss {\tau}
\def \tn {\lambda} 

\newpage
\tableofcontents

\newpage
\section{Introduction}

The critical  first-quantized 
  string  is described, in an appropriate gauge,   by  an effectively Gaussian   path integral.  
This is not  so for a membrane which has a highly  non-linear 
and formally non-renormalizable  3d action. 
While the existence  of a  consistent quantum theory of bosonic membranes may be in doubt, this may not be so 
 for the  11d supermembrane   or M2 brane 
 \ci{Bergshoeff:1987cm,Bergshoeff:1987qx,deWit:1988wri,Duff:1996zn,Nicolai:1998ic}.
 
 This  may be  true, in particular,  for  the supermembrane   in the maximally supersymmetric AdS$_4 \times S^7$
 (or  AdS$_7 \times S^4$)    background 
 \ci{Bergshoeff:1988uc,Duff:1989ez,deWit:1998yu,Pasti:1998tc,Claus:1998fh}
 and its orbifold  AdS$_4 \times S^7/\mathbb Z_k$. 
 M-theory in  AdS$_4 \times S^7/\mathbb Z_k$    should  be  dual to the ${\cal N} =6$
supersymmetric 3d $U_k(N) \times U_{-k}(N)$  Chern-Simons matter (ABJM) theory
 \ci{Aharony:2008ug,Bagger:2012jb}.
Recent   work  \ci{Giombi:2023vzu,Beccaria:2023ujc}   provided
a remarkable evidence   that direct  semiclassical 
quantization of the M2 brane in AdS$_4 \times S^7/\mathbb Z_k$  background 
reproduces the results of   large $N$  localization 
computations  \ci{Drukker:2010nc,Drukker:2011zy,Klemm:2012ii,Hatsuda:2012dt}   
of the  ${\ha}$-BPS Wilson loop and instanton contributions to free energy 
in the ABJM  gauge theory.

While the M2 brane action is  highly non-linear, expanded near a classical  solution with non-degenerate induced 
3d metric  it can be quantized in a static gauge 
with the 1-loop  result being  UV finite (containing no logarithmic  divergences)
\cite{Duff:1987cs,Bergshoeff:1987qx,Mezincescu:1987kj,Forste:1999yj,Drukker:2020swu,Giombi:2023vzu,Beccaria:2023ujc}.
As the $1/N$ expansion of the  localization results  on the gauge theory side 
 have the form of an  expansion 
in the  inverse of  the effective  M2  brane tension  ${\rm T}_2 =\tfrac{\sqrt{ 2k}}{\pi} \sqrt{N}$, 
this   suggests that the matching with the  1-loop 
 M2 brane computations \ci{Giombi:2023vzu,Beccaria:2023ujc}  should, in fact, 
  extend  also to 2-loop   and higher  orders.
  
   This requires
  the corresponding quantum M2   brane theory to    be UV finite despite its apparent non-renormalizability. 
  This  may somehow   happen due to high degree of  underlying   supersymmetry   and possibly to other 
   hidden symmetries of the M2  brane theory in AdS$_4 \times S^7$  background that remain to  be uncovered. 
   
  The M2 brane   action in  11d   background is formally related to the type IIA  string in the corresponding 10d    background   by a double   dimensional reduction  \ci{Duff:1987bx,Achucarro:1989dd}.
    Considering M2  brane world volume  of   topology 
  $\Sigma^2 \times S^1$ 
    and expanding  3d  fields   in Fourier modes 
  in $S^1$  coordinate  one gets an  effective  2d string   action on $\Sigma^2$ 
  coupled to  an  infinite tower  of massive     2d  fields. 
   Choosing a static gauge in the M2 brane action one gets a static  gauge Nambu-Goto    action for the  massless    transverse string modes 
  coupled to a tower of the massive   ``Kaluza-Klein"  2d modes. 
  This  ``effective   string"   2d action is essentially  equivalent to 
   the original M2  brane   action   and thus   may inherit  its   hidden  symmetries.
     
     \
     
  With this  motivation  in mind,  here we  address  the 
   question  about   possible   hidden symmetries 
   in  the simplest context of a bosonic membrane in flat  $\RR^{1, D-1}$  space-time 
    expanded   near a $\RR \times S^1$  cylindrical   vacuum which is the analog of 
    the infinite  straight  string  in the bosonic string theory.
    
 We will focus on   the resulting 2d S-matrix % (at tree and 1-loop level), 
 comparing it 
 with the one found in the   string theory  limit   which  corresponds to the case when the radius
 $\R$ of $S^1$  is sent to zero, i.e. when all  massive KK  2d states  decouple. 
 In the opposite limit $\R\to \infty$ we should recover the S-matrix on plane $\RR^{2}$   membrane.

Let us first recall  some results of past work on 2d S-matrix  in the   infinite    bosonic string 
vacuum. 
The tree-level and 1-loop contributions to the scattering of 
four  massless  modes    representing the $\hat D\equiv D-2$ transverse string coordinates 
 of  the NG string in the  static gauge was computed in \cite{Dubovsky:2012sh}.
 The S-matrix was found  to be given  simply  by a scalar (CDD) phase factor 
 thus  representing  an integrable theory. 
  It was shown in    \cite{Dubovsky:2012wk}    that using this S-matrix in the thermodynamic Bethe Ansatz 
    one reproduces    the expected free  bosonic string spectrum. 
   This   S-matrix   was  studied  at  higher  loop orders  \cite{Conkey:2016qju}  where particular  counterterms  are  required 
    to  cancel UV divergences   and also to  preserve  the  integrability.
The reason why  this  S-matrix is  given simply by a pure phase factor was further elucidated in \cite{Cavaglia:2016oda} by  observing  
that the NG action in a static gauge can be viewed as the $T\bar{T}$ deformation \cite{Zamolodchikov:2004ce}
 of a theory of free massless bosons.

One may   wonder if the integrability property of the NG string   may generalize  to a 
2d theory containing a special   infinite set of  massive 2d modes. Our aim   will be to 
address this question  for the   theory  obtained  from the
 bosonic membrane action  in the static gauge  with $S^1_\R$  compact direction 
  viewed as an effective 2d theory with   a tower of states with masses  $m^2_{n}=\frac{n^2}{\R^2}$, $n=0, \pm 1, ...$. 
  As mentioned above, this  spectrum 
    corresponds to the  expansion near the  cylindrical $\RR \times S^1$  membrane 
   vacuum. 
   
  We will  discuss  the corresponding tree-level scattering amplitudes and  conclude  that this   model  involving massive fields in addition to massless ones 
  is no longer integrable. 
  We will also compute 
  the 1-loop correction to the   scattering  amplitude  of 4  massless  particles  with 
  all  massless and massive modes  running in the loop.
   We will find that as in the examples  considered in \ci{Giombi:2023vzu,Beccaria:2023ujc}
   this 1-loop amplitude is  UV finite in  analytic regularization 
     as  appropriate for a 3d theory. It is    
   also IR finite as all massless modes  have  only derivative couplings  like in the NG string  case.
  The 1-loop amplitude   has a  non-trivial dependence on the  kinematic variables  (and the mass  scale or the 
   radius $\R$),  
  characteristic of a non-integrable  theory.

  For comparison,  we will also  consider a 2d  model with the same free massless  plus massive tower   spectrum 
  but with interactions  defined   by  the   $T\bar{T}$ deformation. 
  At the massless 
  $n=0$   level the resulting action  will  be given by the same NG   action  but 
  interaction   vertices   involving massive fields
    will be different from  those   in  the compactified   membrane   action  
  (and, in fact,   will not  follow  from  any  local 3d  action compactified on $S^1$). 
  As this is a $T\bar{T}$ deformation of a  free theory, 
   it should be integrable, and we shall verify this  by  computing  tree-level  and 1-loop  scattering amplitudes.

  \

  This  paper is organized as follows. In section 2  we shall review in detail  the tree-level and 1-loop 
  computations  in the NG   model near  infinitely long string  vacuum, verifying  that the S-matrix  admits  a pure-phase representation.
  
     Section 3  will be   dedicated to a similar  analysis in the  compactified membrane theory. We shall compute 1-loop  amplitudes  for   scattering  of 4 massless modes 
  using dimensional regularization near $d=2$   and Riemann $\zeta$-function regularization to sum over  KK  modes. 
  We will also present (in section 3.4)  the expression for the 1-loop   amplitude in the uncompactified membrane model
  found   by starting directly  from the 3d  action. 
  
  The 2d model obtained by $T\bar T$ deformation  of  the same  free  massless  plus massive spectrum 
   will be  studied  in Section 4.  
   Some  open problems   will be  mentioned in section 5. 
   Appendices will contain some  basic definitions, useful 1-loop integrals, 
   comments on 6-point tree-level amplitudes and details of  computations in section 3.3.

   %%%%%%%%%%%%%%%%%%%%%%%%
   
\section{S-matrix on Nambu-Goto string and its integrability}
\la{s2}

Let us start with a  review  of the perturbative S-matrix of the  Nambu-Goto  string action
in the long  string   vacuum  \cite{Dubovsky:2012sh}. 
Fixing  static   gauge 
one may   study   scattering  of the 2d   massless fields representing the transverse  string coordinates. 
One finds that  there is no particle production or annihilation at tree-level, 
which is compatible with the classical integrability of this 2d  model. The  loop-corrected 
S-matrix is   given  by a pure phase  
provided appropriate counterterms are added, which 
 is consistent with  known free string spectrum  \cite{Dubovsky:2012wk}.\foot{Also, the NG string action in  the static gauge can be obtained from the $T\bar{T}$ deformation of the free action for a set of $D-2$  massless fields \cite{Cavaglia:2016oda,Bonelli:2018kik}  and therefore the  S-matrix  should   be 
 given  simply  by a CDD factor, dressing the trivial S-matrix of free massless scalars. Note that  certain 
 %When computing the 1-loop S-matrix from the classical Lagrangian, 
 counterterms are to be added to the classical Lagrangian to  maintain  integrability
 at  loop level  and  thus to match with the expected CDD factor.
 Such counterterms were 
   explicitly worked out for the $T\bar{T}$ deformation of a  single 
   massive scalar field in \cite{Rosenhaus:2019utc}.}
   
   Below we shall describe the  computation of the 1-loop  NG string S-matrix in some detail to prepare  for 
   the   analysis of the  membrane case in the next section.
 
Let us start with a brane with 1 time and $d-1$ space world-volume directions moving in a $D$-dimensional target space-time.
We will be interested in the cases of the string with $d=2$ and the membrane with $d=3$. 
The corresponding Dirac \cite{Dirac:1962iy} or Nambu-Goto action is
\begin{equation}
S = -T_{d-1}  \int d^d \sigma \sqrt{-\text{det} \, \gamma}~, \qquad \qquad  \gamma_{\alpha \beta} = \eta_{\mu \nu} \partial_\alpha X^\mu \partial_\beta X^\nu~ \ ,  \la{21}
\end{equation}
where   $\alpha,\beta=0,1,\dots d-1$   and  $\mu,\nu =0,1,\dots,D-1$.
We shall   consider the expansion of \rf{21} 
near   an  infinite flat string or membrane   vacuum   so that we can fix  the    static gauge 
$X^\a=\sigma^\a $. We shall  label the remaining transverse coordinates as $X^j$   so that  the 
induced metric is 
\begin{equation}
\gamma_{\alpha \beta} = \eta_{\alpha \beta} + \partial_\alpha X^j \partial_\beta X^j~,\qquad 
\ \   j=1, ..., \hD \ , \ \ \ \ \ \ \hD \equiv D-d \ . \la{23}
\end{equation}
Then expanded in derivatives of $X^j$   the  action \rf{21} takes the form 
\begin{align} \label{22}
 S &= - T_{d-1} \int d^d \sigma \Big(1 + \frac{1}{2} \partial_\alpha X^j \partial^\alpha X^j  + \mathcal L_4 +
\mathcal L_6 +  \dots  \Big)~,\\
 \mathcal L_4 &= \frac{1}{4} \Big( c_2 \partial_\alpha X^j \partial^\alpha X^j \partial_\beta X^k \partial^\beta X^k + c_3 \partial_\alpha X^j \partial^\alpha X^k \partial_\beta X^j \partial^\beta X^k \Big)~,\quad \ \ \ 
 c_2=\frac{1}{2}, \quad c_3=-1 \ . \la{24}
\end{align}
Let us define 
\begin{equation}\la{25}
\mathcal X_{\alpha \beta} \equiv  \partial_\alpha X^j \partial_\beta X^j, \qquad\ \ \ \  \JJ_k \equiv  \Tr[\mathcal X^k],
\end{equation} 
so that 
\begin{align}\la{27} 
\mathcal L_2 =  &\frac{1}{2} \JJ_1, \qquad \mathcal L_4 = \frac{1}{4} \Big( c_2 \JJ_1^2 + c_3 \JJ_2 \Big), \qquad 
\mathcal L_6 = \frac{1}{6} \Big(c_4 \JJ_1^3 +c_5 \JJ_1 \JJ_2 + c_6 \JJ_3 \Big), \ \ \ ...\\  \la{26}
&  \qquad   \qquad   c_4 = \frac{1}{8}, \ \qquad   c_5 = -\frac{3}{4}, \ \qquad  c_6=1 \ . 
\end{align}
\iffa with coefficients
\begin{equation} \label{26}
c_4 = \frac{1}{8}, \qquad c_5 = -\frac{3}{4}, \qquad c_6=1.
\end{equation}
\fi
Notice that  in the string case when  $d=2$ only $\JJ_1$ and $\JJ_2$ are independent 
invariants; for example,    $\JJ_3= - \frac{1}{2} \JJ_1^3 + \frac{3}{2} \JJ_1 \JJ_2$. 
Below   we shall sometimes   keep the   values of the coefficients $c_n$ in \rf{24},\rf{26} 
arbitrary to emphasize simplifications that happen when they take their  ``Nambu-Goto"  values. 

%%%%%%%%%%%%%%
\subsection{Tree-level 4-point amplitude}
\la{s12}

Starting with \rf{22} one  can compute the scattering amplitudes of the massless  fields $X^j$. 
From the expression of $\mathcal L_4$  in \rf{24} we deduce that there are two types of Feynman diagrams contributing to the four-point amplitude, 
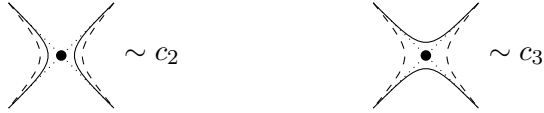
\begin{center}
\begin{tikzpicture}[scale=0.7]
\draw[-] (-1,1) .. controls (0,0) .. (-1,-1) ; 
\draw[-] (1,1) .. controls (0,0) .. (1,-1) ; 
\draw[dashed] (-1,1) .. controls (-0.2,0) .. (-1,-1) ; 
\draw[dashed] (1,1) .. controls (0.2,0) .. (1,-1) ; 
\fill (0,0)  circle (0.1);
\draw[dotted] (-1,1) -- (1,-1);
\draw[dotted] (-1,-1) -- (1,1);
\node[anchor=west] at (1,0) {$\sim c_2$};
\end{tikzpicture}  \qquad  \qquad  \qquad 
\begin{tikzpicture}[scale=0.7]
\draw[-] (-1,1) .. controls (0,0) .. (1,1) ; 
\draw[-] (-1,-1) .. controls (0,0) .. (1,-1) ; 
\draw[dashed] (-1,1) .. controls (-0.2,0) .. (-1,-1) ; 
\draw[dashed] (1,1) .. controls (0.2,0) .. (1,-1) ; 
\fill (0,0)  circle (0.1);
\draw[dotted] (-1,1) -- (1,-1);
\draw[dotted] (-1,-1) -- (1,1);
\node[anchor=west] at (1,0) {$\sim c_3$};
\end{tikzpicture} 
\end{center}
The solid line here represents  contraction of momenta, while the dashed lines correspond to a contraction of indices of the fields. 
Let us label the two incoming particles with indices $i,j$ and momenta $p_1,p_2$ and the  two outgoing particles with indices $k,l$ and momenta $p_3,p_4$.
The amplitude takes the  following schematic form (we omit the standard momentum conservation delta-function, \rf{a15},\rf{a7})
\begin{equation} \label{222}
\mathcal M_{ij,kl} = A\, \delta_{ij} \delta_{kl} + B\, \delta_{ik} \delta_{jl} + C\, \delta_{il} \delta_{jk}  %\Big)
% \delta^{(d)}(p_1 + p_2 - p_3 - p_4)
~,
\end{equation}
where $A$ is the annihilation, $B$  the transmission and $C$  the reflection  parts. 
Their tree-level expressions  following from \rf{24} 
are\foot{\la{f1} After rescaling $X^i\to \tfrac{1}{ \sqrt{T}} X^i$  in \rf{22} 
where $T\equiv T_1$ is the   string tension 
  the factors of  $T^{-1}$ 
or effective $\hbar$  appear  in the quartic, etc., interaction   vertices   and thus
in   the corresponding scattering amplitudes. 
We will  not always  explicitly include them below  as they can be restored in the final expressions.}
%v4 overall signs
\begin{align} 
A^{(0)}[p_1,p_2,p_3,p_4] &=-\Big[ 2 c_2 (p_1 \cdot p_2) (p_3 \cdot p_4)+ c_3 (p_1 \cdot p_4)(p_2 \cdot p_3) + c_3 (p_1 \cdot p_3)(p_2 \cdot p_4)\Big]~,\no  \\
B^{(0)}[p_1,p_2,p_3,p_4] &= -\Big[2 c_2 (p_1 \cdot p_3)(p_2 \cdot p_4) + c_3 (p_1 \cdot p_2)(p_3 \cdot p_4) + c_3 (p_1 \cdot p_4)(p_2 \cdot p_3)\Big]~, \la{210}\\
C^{(0)}[p_1,p_2,p_3,p_4] &=-\Big[ 2c_2 (p_1 \cdot p_4)(p_2 \cdot p_3) + c_3 (p_1 \cdot p_2)(p_3 \cdot p_4) + c_3 (p_1 \cdot p_3)(p_2 \cdot p_4)\Big]~.\no 
\end{align}
Writing  these   in terms of  the Mandelstam variables (see Appendix \ref{a1})
taking into account that 
here all particles are massless, i.e. $s+t+u=0$,  we get 
\begin{align}\la{28}
%v4 overall signs
 A^{(0)}= - \frac{1}{4} (2 c_2 + c_3) s^2 &+ \frac{1}{2} c_3 t u~, \qquad B^{(0)}= -\frac{1}{4} (2 c_2 + c_3) t^2 +\frac{1}{2} c_3 s u~, \no\\
 & C^{(0)}= - \frac{1}{4} (2 c_2 + c_3) u^2 +\frac{1}{2} c_3 st~.
\end{align}
For the  values  of the coefficients in \rf{24} this becomes 
\begin{equation}\la{29}
%v4 overall signs
A^{(0)} = -\frac{1}{2} tu~,\ \ \  \qquad B^{(0)} = -\frac{1}{2} su~,\ \ \  \qquad C^{(0)}= -\frac{1}{2} st~.
\end{equation}
Specifying to  $d=2$  we may use that in this case the   kinematical constraints 
 imply  that   for massless  particles  $stu=0$ or   $tu=0$ if we assume that $s\not=0$ (see \rf{a12})\foot{For example, we may use a Lorentz transformation 
to go to  the center of mass frame   where  $\vec{p}_1=-\vec{p}_2\not=0$, i.e. 
 choosing to keep $s\not=0$ (here $\vec p$   is  spatial  component of  momentum, see Appendix \rf{a1} for notation). 
 Then by the momentum conservation 
$\vec{p}_3 = - \vec{p}_4$   and the energy conservation   gives 
 $|\vec p_1| = |\vec p_3|$   or  $tu=0$.}
\be   s t u =0 \  \ \to \ \    tu=0 \ , \ \ \  \ s\not=0 \ .  \la{211}\ee
 With   this choice  of the kinematics we  get $A^{(0)}=0$ in \rf{29}. Also, 
  $B^{(0)}=0$ for $u=0$ and $C^{(0)}=0$ for $t=0$.
   For the latter case the tree-level S-matrix \rf{222} 
  contains only the  $B\zz \delta_{ik} \delta_{jl}$ term    which is proportional to the  unit operator in the S-matrix 
  (cf. \rf{a44},\rf{a14}).
   Being proportional to the identity, the tree-level S-matrix automatically satisfies the  Yang-Baxter equation
   \rf{a9},\rf{a10}.

\subsection{One-loop contribution to the 4-point amplitude}

Using the 4-vertex  in \rf{24} we can build three 1-loop ``bubble'' diagrams  in Figure \ref{fig:bubble}.
The total 1-loop amplitude, given by the sum of these three contributions,  takes the same  form  as \rf{222}
(cf. footnote \ref{f1})
\begin{align} \label{2133}
\mathcal M^{(1)}_{ij,kl} &= %\Big(
A^{(1)} \delta_{ij} \delta_{kl} + B^{(1)} \delta_{ik} \delta_{jl} + C^{(1)} \delta_{il} \delta_{jk}
%\Big) \delta^{(d)}(p_1+p_2-p_3-p_4)
~.
\end{align}
Each amplitude $A^{(1)}$, $B^{(1)}$ and $C^{(1)}$ have a contribution coming from the $s$-channel, $t$-channel and $u$-channel diagrams, i.e. 
\be  A^{(1)}=A_s^{(1)} + A_t^{(1)} + A_u^{(1)}\ , \qquad 
 \label{213}
B^{(1)} = \left. A^{(1)} \right|_{s \leftrightarrow t}~, \qquad C^{(1)} = \left. A^{(1)} \right|_{s \leftrightarrow u}~.
\end{equation}
%%%%%%%%%%%%%%%%%%%%%%%%
\begin{figure}
\begin{tikzpicture}[scale=0.7]
\begin{scope}[decoration={
    markings,
    mark=at position 0.5 with {\arrow{latex}}}
    ] 
\draw[postaction={decorate}] (-2,2) node[left] {$(p_1,i)$} -- (-1,0);
\draw[postaction={decorate}] (-1,0) to [out=60,in=120] node[midway, above] {$(p,a)$}  (1,0);
\draw[postaction={decorate}](1,0) -- (2,2) node[right]{$(p_3,k)$};
\draw[postaction={decorate}] (-2,-2) node[left] {$(p_2,j)$}--(-1,0);
\draw[postaction={decorate}] (-1,0) to [out=-60,in=-120]  node[midway, below] {$(q,b)$} (1,0);
\draw[postaction={decorate}] (1,0) -- (2,-2) node[right] {$(p_4,l)$};
\end{scope}
\fill (-1,0) circle (0.1); 
\fill (1,0) circle (0.1); 
\draw[] (0,-3) node[] {$s$-channel};
\end{tikzpicture} \qquad
\begin{tikzpicture}[scale=0.7]
\begin{scope}[decoration={
    markings,
    mark=at position 0.5 with {\arrow{latex}}}
    ] 
\draw[postaction={decorate}] (-1,2) node[left] {$(p_1,i)$}--(0,1);
\draw[postaction={decorate}] (0,1) -- (1,2) node[right] {$(p_3,k)$};
\draw[postaction={decorate}] (0,1) to [out=-150,in=150]  node[midway, left] {$(p,a)$} (0,-1);
\draw[postaction={decorate}] (0,1) to [out=-30,in=30] node[midway, right] {$(q,b)$}  (0,-1);
\draw[postaction={decorate}] (0,-1) -- (1,-2) node[right]{$(p_4,l)$};
\draw[postaction={decorate}] (-1,-2) node[left] {$(p_2,j)$}--(0,-1);
\end{scope}
\fill (0,1) circle (0.1); 
\fill (0,-1) circle (0.1); 
\draw[] (0,-3) node[] {$t$-channel};
\end{tikzpicture} \qquad
\begin{tikzpicture}[scale=0.7]

\draw[-latex] (0,1) .. controls (2,2) .. (2.5,-2) node[right] {$(p_4,l)$};
\draw[-latex]  (0,-1) .. controls (2,-2) ..  (2.5,2) node[right]{$(p_3,k)$};
\begin{scope}[decoration={
    markings,
    mark=at position 0.5 with {\arrow{latex}}}
    ] 
\draw[postaction={decorate}] (-1,2) node[left] {$(p_1,i)$}--(0,1);
\draw[postaction={decorate}] (0,1) to [out=-150,in=150]  node[midway, left] {$(p,a)$} (0,-1);
\draw[postaction={decorate}] (0,1) to [out=-30,in=30] node[midway, right] {$(q,b)$}  (0,-1);
\draw[postaction={decorate}] (-1,-2) node[left] {$(p_2,j)$}--(0,-1);
\end{scope}
\fill (0,1) circle (0.1); 
\fill (0,-1) circle (0.1);
\draw[] (0,-3) node[] {$u$-channel}; 
\end{tikzpicture} 
\caption{\small The three ``bubble'' diagrams contributing to the 1-loop 4-point amplitude. 
The arrows denote the flow of momentum: the particles $p_1$ and $p_2$ are  incoming, while the particles $p_3$ and $p_4$ are outgoing. 
By momentum conservation we have $q=p_1+p_2-p$ for the first diagram ($s$-channel), $q=p_1-p_3-p$ for the second diagram ($t$-channel) and $q=p_1-p_4-p$ for the third diagram ($u$-channel).}
\label{fig:bubble}
\end{figure}
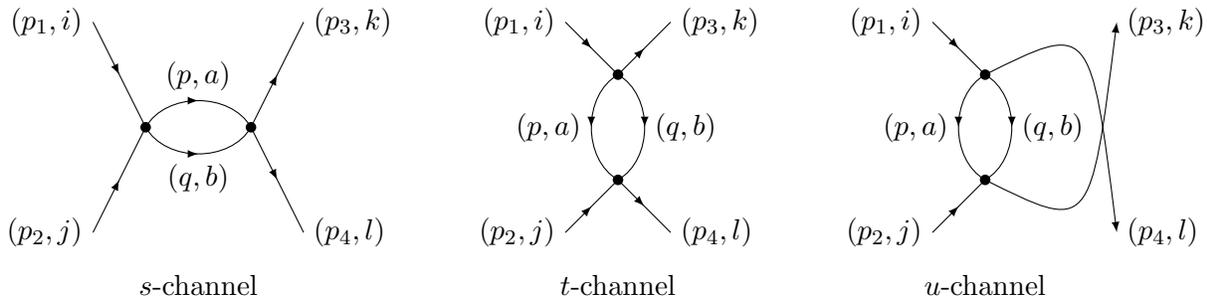 
%%%%%%%%%%%%%%%%%%%
The  corresponding loop integrals diverge  for  $d=2$. 
To define them we will use dimensional regularisation   setting 
 $d=2-2\epsilon, \  \  \epsilon \to0$ (see   Appendix \ref{a2} for some standard momentum integrals). 

Using Feynman parametrization,  the  $s$-channel  amplitude $A_s^{(1)}$ in \rf{213} 
 may be written as\foot{Note that $\varepsilon\to 0 $ in the 
Feynman propagator should not be confused with $\epsilon$ of the dimensional regularisation.
The $-i \ve$ term in $\Delta_s$ translates into $s \to s + i\ve$. We shall  not explicitly 
indicate this shift  below (same will apply also for $t$ and $u$).} 
\begin{align} 
\label{215}
A_s^{(1)} =& \frac{1}{2i} \int \frac{d^d p}{(2 \pi)^d} \frac{N_s}{(p^2-i \varepsilon) [(p_1+p_2-p)^2- i \varepsilon]} = \frac{1}{2i} \int_0^1 d x  \int \frac{d^d l}{(2 \pi)^d} \frac{N_s}{(l^2 + \Delta_s)^2},\\
& \qquad \qquad  l = p-x(p_1+p_2), \qquad\qquad  \Delta_s = -x(1-x) s- i \varepsilon,\la{512}
\end{align}
where  $\ha$  is the symmetry factor   of the diagram 
(exchange of   two legs in the loop) and  %.
\begin{align}
N_s &= \hat{D} A\zz[p_1,p_2,p,q] A\zz[p,q,p_3,p_4] + A\zz[p_1,p_2,p,q]\big(B\zz[p,q,p_3,p_4] + C\zz[p,q,p_3,p_4]\big)\no  \\
&\qquad  +\big(B\zz[p_1,p_2,p,q] + C\zz[p_1,p_2,p,q]\big) A\zz[p,q,p_3,p_4]\ , \qquad  q=p_1+p_2-p \ .  \label{219}
\end{align}
Here  the factor $\hat{D}=D-2$ of  the number of transverse fields  
comes from a diagram with a  sum   over the target space indices in the loop
   and $A\zz, B\zz, C\zz$ were defined in 
 \rf{210}. To simplify the expressions   below we  will assume that 
  the coefficients $c_2$ and $c_3$ take  their NG  values  in \rf{24}.

It is useful to formally expand $N_s$ \rf{219}  in powers of loop momentum  $l$  as 
\begin{align} \label{288}
N_s &= N_0 + N_2 l^2 + N_{\mu \nu} l^\mu l^\nu  + N_4 (l^2)^2+N_{\mu \nu \rho \sigma} l^\mu l^\nu l^\rho l^\sigma + M_{\mu \nu} l^2 l^\mu l^\nu \ , \\
N_0 &= 0~,  \quad N_2 = -s^3 x(1-x)~,  \quad N_4 = \frac{1}{4} (\hat{D}-4)s^2~, \quad
N_{\mu \nu} = -2 s^2 x(1-x)  (p_{1,\mu} p_{2,\nu}+ p_{3,\mu}p_{4,\nu} ),\no  \\
M_{\mu \nu} &= (\hat{D}-2) s \big(p_{1,\mu} p_{2,\nu}
+p_{3,\mu} p_{4,\nu}\big)~, \qquad 
N_{\mu \nu \rho \sigma} = 4 \hat{D} p_{1,\mu} p_{2,\nu} p_{3,\rho} p_{4,\sigma}~.
\la{2199} \end{align}
Performing  the integral over $l$ in   \rf{215}    we get (cf.  Appendix \ref{a2})  
\begin{equation} \la{218}
A_s^{(1)} = \frac{1}{2} \frac{1}{(4 \pi)^{\frac{d}{2}}} \int_0^1 dx \, \Big[ \BB_0\,  \Gamma(2-\tfrac{d}{2}) \, (\Delta_s)^{\frac{d}{2}-2} + \BB_2\,  \Gamma(1-\tfrac{d}{2}) \,( \Delta_s)^{\frac{d}{2}-1}+\BB_4\,  \Gamma(-\tfrac{d}{2}) \, (\Delta_s)^{\frac{d}{2}}  \Big]\ , 
\end{equation}
where 
\begin{equation} \label{eq:beta}
\begin{aligned}
\BB_0 &= N_0 ~, \qquad \BB_2 = \frac{d}{2} N_2 + \frac{1}{2} N_{\mu \nu} \eta^{\mu \nu}  ~, \\
 \BB_4 &= \frac{d}{2} \Big( 1+\frac{d}{2} \Big)N_4+ \frac{1}{4} N_{\mu \nu \rho \sigma} \Big(\eta^{\mu \nu} \eta^{\rho \sigma} + \eta^{\mu \rho} \eta^{\nu \sigma} + \eta^{\mu \sigma} \eta^{\nu \rho} \Big) + \frac{1}{2} \Big(1+\frac{d}{2}\Big) M_{\mu \nu} \eta^{\mu \nu} ~.
 \end{aligned}
\end{equation}
Using  standard $\Gamma$-function relations   implying, e.g., that  
\begin{equation}\la{2211}
\tfrac{d}{2} \big(1+\tfrac{d}{2}\big) \Gamma(-\tfrac{d}{2}) \, ({\Delta_s})^{\frac{d}{2}} =( \Delta_s)^2 \Gamma(2-\tfrac{d}{2}) \, (\Delta_s)^{\frac{d}{2}-2} -2 \Delta_s \Gamma(1-\tfrac{d}{2})\,  ({\Delta_s})^{\frac{d}{2}-1}, 
\end{equation}
we can  rewrite \rf{218} as 
\begin{align}
\la{221}
A_s^{(1)} &= \frac{1}{2} \frac{1}{(4 \pi)^{\frac{d}{2}}} \int_0^1 dx \, \Big[ \GG_0\,  \Gamma(2-\tfrac{d}{2})\,  (\Delta_s)^{\frac{d}{2}-2} + \GG_2\, \Gamma(1-\tfrac{d}{2})\,  (\Delta_s)^{\frac{d}{2}-1}+\GG_4\, \Gamma(-\tfrac{d}{2})\, ( \Delta_s)^{\frac{d}{2}}  \Big] \ , 
\\  &
\label{223}
\GG_0 = N_0 - \Delta_s N_2 +\Delta_s^2 N_4 \ , \qquad \qquad 
\GG_2 = N_2 + \frac{1}{2} N_{\mu \nu} \eta^{\mu \nu} - 2 \Delta_s N_4 - \frac{1}{2} \Delta_s M_{\mu \nu} \eta^{\mu \nu} \ , \\
& \GG_4 = \frac{1}{4} N_{\mu \nu \rho \delta} \Big(\eta^{\mu \nu} \eta^{\rho \sigma} + \eta^{\mu \rho} \eta^{\nu \sigma} + \eta^{\mu \sigma} \eta^{\nu \rho} \Big)  + \frac{1}{2} M_{\mu \nu} \eta^{\mu \nu}\ . \la{2233}
\end{align}
The  values of the  coefficient functions $\g_0, \g_2, \g_4$  are found from \rf{2199}
 not to depend  on $d$  (cf.  \rf{512})
\begin{equation}\la{2255}
\GG_0 = \frac{1}{4}(\hat{D}-8) \, s^2 \Delta_s^2\ , \qquad\qquad \GG_2 = s^2\,  \Delta_s\ , \qquad\qquad  \GG_4 = s^2 -\frac{1}{2}\,\hat{D}\,  t u  \ .
\end{equation}
Setting $d=2-2 \epsilon$ it is then straightforward  to isolate  the  divergent and finite part of the amplitude 
 in the $\epsilon \to 0$ limit:
\begin{align}
A_s^{(1)} &= \frac{1}{8 \pi} \int_0^1 dx \, \Big( \frac{1}{\epsilon} - \gamma + \ln 4 \pi \Big) \big( \GG_2 -\Delta_s \GG_4 \big)\no  \\
&\ \ + \frac{1}{8 \pi} \int_0^1 dx \, \Big[\GG_0\,  {\Delta^{-1}_s} - \GG_2\,  \ln \Delta_s + \GG_4\,  \Delta_s  \big(\ln \Delta_s-1\big)  \Big]\ . \la{224}
\end{align}
Similar  expressions   are found in the $t$ and $u$ channels,  where,  e.g., 
\begin{align} 
\label{eq:At1}
A_t^{(1)} & = \frac{1}{2i} \int \frac{d^d p}{(2 \pi)^d} \frac{N_t}{(p^2-i\ve) [ (p_1-p_3-p)^2-i\ve]
} = \frac{1}{2i} \int_0^1 d x  \int \frac{d^d l}{(2 \pi)^d} \frac{N_t}{(l^2 + \Delta_t)^2}\ ,\\
& l = p-(p_1-p_3), \qquad \Delta_t = -x(1-x) t - i \ve\no  \ ,\\
N_t &= \Big(A\zz[p_1,-p,p_3,q] \, A\zz[p,p_2,-q,p_4] + C\zz[p_1,-p,p_3,q]\,  C\zz[p,p_2,-q,p_4]\Big)\Big|_{  q=p_1-p_3-p   }\ , \no \end{align}
and thus we get  \rf{224} where  now 
$\GG_0 = 2 t^2 \Delta_t^2,  \ 
\GG_2 = -3 t^2 \Delta_t +\frac{1}{2} s t^2, \ 
 \GG_4 =0$.
Performing the  integrals over the Feynman parameter $x$ gives  
\begin{align}\la{226}
A_s^{(1)} 
= & -\frac{1}{96 \pi} \Big( \frac{1}{\epsilon} - \gamma + \ln 4 \pi \Big)  \hat{D} s t u   - \frac{1}{192 \pi} \Big[ (\hat{D}-12) s^3 + \frac{16}{3} \hat{D} \, s t u - 2 \hat{D}  \ln (-s) \, s t u \Big], \\
\la{230}
A_t^{(1)} = & -\frac{1}{16 \pi} \Big( \frac{1}{\epsilon} - \gamma + \ln 4 \pi \Big)  t^2 u  +\frac{1}{192 \pi} \Big[ 12 t^3 +
 24 t^2 s - 12 t^2 u \ln \big(\frac{1}{-t}\big) \Big]\  , \\ 
 \la{231}
A_u^{(1)} = & -\frac{1}{16 \pi} \Big( \frac{1}{\epsilon} - \gamma + \ln 4 \pi \Big) t u^2  +\frac{1}{192 \pi} \Big[ 12 u^3 +24 u^2 s - 12 u^2 t \ln \big(\frac{1}{-u}\big) \Big]\ . 
\end{align}
Adding \rf{226},\rf{230} and \rf{231}    together  we get for the
   divergent and finite parts  of  $A^{(1)}$ 
   \begin{align}
\label{2311}
&  \ \ \ \ \qquad  \ \ A^{(1)}_{\epsilon}  = - \frac{\hat{D}-6}{96 \pi} \Big( \frac{1}{\epsilon}-\gamma + \ln 4 \pi \Big) s t u~,\\
%tribution\begin{equation}
A^{(1)}_f &= -\frac{1}{192 \pi} \Big[ (\hat{D}-24) s^3 + \Big( \frac{16}{3} \hat{D} + 12 - 2 (\hat{D}-6) \log \frac{-s}{\mu^2} \Big) s t u + 12 \Big( t \log \frac{s}{t} + u \log \frac{s}{u} \Big) t u \Big].\no
\end{align}
Taking into account that in the string case $\hat{D}=D-2$
this is the same expression as found originally in    \cite{Dubovsky:2012sh}.
The divergent part  in \rf{2311}    is proportional  to $stu$ and thus vanishes in $d=2$  due to  \rf{211}
(thus  the dependence on  renormalization scale  $\mu$   drops out). 
The same applies to the   $B\zo$ and $C\zo$  amplitudes related to $A\zo$ as in \rf{213}.\foot{As 
$stu= \OO(d-2)$  this is  an ``evanescent"     contribution 
% there may be   an  ``evanescent"   finite    contribution  but it 
 % can be cancelled by a  local counterterm 
(cf. footnote \ref{f4}   and ref.   \cite{Dubovsky:2012sh}).
%v4
This  UV  divergent term corresponds  to $ \int d^d\sigma\,  \sqrt{-\gamma}\,  R $   divergence   which 
is topological in 2d; its   coefficient $\hat{D}-6 = D-8$ matches  the one  found by quantizing Polyakov string action 
on a general curved 2d background \ci{Fradkin:1981dd}.
 }
%%%%%%%%%%%
\iffa
The amplitude in \eqref{2311} is proportional to $stu$ and hence invariant under exchanges of the Mandelstam variables, so that $A^{(1)}_\epsilon=B^{(1)}_{\epsilon} =C^{(1)}_{\epsilon}$. Moreover, in 2d, the kinematic constraint imposes $tu=0$. Due to the $\frac{1}{\epsilon}$ prefactor, this term therefore is not divergent, but rather contributes to the finite part of the amplitude.
\fi 
%%%%%%%%%%%%%

 Let us  note that there is also another  1-loop  diagram
   that could   contribute to the 4-point amplitude   --  the  tadpole  with 6-vertex from $\mathcal L_6$ in \rf{27} 
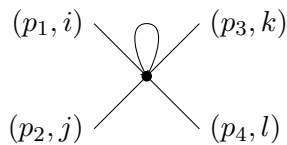
\begin{center}
\begin{tikzpicture}[scale=0.7]
\draw[-] (-1,-1) node[left] {$(p_2,j)$}--(1,1) node[right] {$(p_3,k)$};
\draw[-] (-1,1) node[left] {$(p_1,i)$}--(1,-1) node[right] {$(p_4,l)$};
\fill (0,0) circle (0.1); 
\draw[->] (0,0) to [out=60,in=0] (0,1) to[out=-180,in=120]  (0,0);
\end{tikzpicture} 
\end{center}
  but it vanishes in dimensional regularization (the vertex   contains powers of momenta
  and the propagator  here is   massless).\foot{Explicitly,  the tadpole     contribution  found   from ${\cal L}_6$  in \rf{27} 
 is 
$ %\begin{equation}
\frac{\Gamma (-1+ \epsilon)}{16 \pi}  \big({-i \varepsilon} \big)^{1-\epsilon} \big[\frac{1}{2}(4 c_4 + 3 c_5 + c_6) s^2 - (c_5+c_6) t u \big].
$, 
Here  the   $\tfrac{1}{\epsilon}$ pole  is  proportional to the effective ``mass" term    $-i \ve$ in the propagator (cf. \rf{215}) 
and thus  vanishes  for $\ve \to 0$.}

The  finite part   of   the 1-loop amplitude is thus given by 
$A^{(1)}_f$  in \rf{2311}   and similar expressions for $B^{(1)}_f$  and $C^{(1)}_f$  obtained by interchanging $s\leftrightarrow t$ and $s\leftrightarrow u$ as in  \rf{213}. 
 The first term in $A^{(1)}_f$ 
 is proportional to $\hat{D}-24=D-26$ and hence vanishes in the  critical dimension of the bosonic string.
 The second term  is proportional to $stu$ and hence vanishes in $d=2$.
  Thus for   $s\not=0$, i.e. $tu=0$
\begin{align}
A^{(1)}_f &=- \frac{\hat{D}-24}{192 \pi} s^3~, \la{232}\\
B^{(1)}_f &= - \frac{1}{192 \pi} \Big[ (\hat{D}-24) t^3 + 12  \Big( s \log \frac{t}{s} + u \log \frac{t}{u} \Big) su \Big]~, \la{233}\\
 C^{(1)}_f &= - \frac{1}{192 \pi} \Big[ (\hat{D}-24) u^3 + 12 \Big( t \log \frac{u}{t} + s \log \frac{u}{s} \Big) st \Big]~.\la{234}
\end{align}
Assuming that the kinematical constraint  $tu=0$   is solved by $t=0$
 and thus $u=-s$    we get\foot{We use that $\log(-s-i \varepsilon) = \log s  - i \pi+O(\varepsilon)$.}
 \begin{align}
A^{(1)}_f =- \frac{\hat{D}-24}{192 \pi} s^3~, \qquad \qquad 
B^{(1)}_f =  \frac{i}{16}  s^3 ~, \qquad\qquad 
 C^{(1)}_f =  \frac{\hat{D}-24}{192 \pi} s^3~.\la{235}
\end{align}
When $\hat D=24$, i.e. $D=26$, the 1-loop amplitude \rf{2133} 
thus contains  only the $B^{(1)}$ term, i.e.  like the tree-level  amplitude \rf{222},\rf{29} it 
is proportional  to the identity and  satisfies the Yang-Baxter equation. 

The same     is true also  when  $\hat{D}=1$ or  $ D=3$, i.e. when  there is just one 
``transverse"  scattering field,  so that the amplitude  in \rf{2133}  is proportional to 
\begin{equation}
 A^{(1)}_f + B^{(1)}_f + C^{(1)}_f = \frac{i}{16} s^3\ . \la{236} 
\end{equation}
For generic  dimension $\hat D=D-2$  the 1-loop S-matrix \rf{235}
does not have a CDD   phase factor form. 
  However, as expected from the relation  of the  static-gauge NG  action  to a $T\bar{T}$ deformation of a theory of free $\hat D$ scalar fields  mentioned above, it should be possible to restore this pure phase structure and consistency with the Yang-Baxter equation    by  adding  contributions of appropriate local  
   counterterms.
At  quartic order in the fields $X^i$   
there exist two  linearly independent local  terms  invariant under 
 $ISO(1,1) \times SO(\hat{D})$   which are  of 6-th order in derivatives
\begin{equation}
\label{237}
\mathcal L_4^{c.t.} = b_1\,  \partial^\beta X^j \partial^\gamma X^k \partial_{\alpha} \partial_\beta X^k \partial^\alpha \partial_\gamma X^j + b_2\,  (\partial_\alpha \partial_\beta X^j \partial^\beta X^j)^2 \ . 
\end{equation}
 The corresponding   tree-level amplitude is given by \rf{222}  with 
\begin{equation}
A^{c.t} = -\ha ( b_1  s t u + b_2 s^3) , \qquad\ \  B^{c.t.} =  -\ha ( b_1  s t u + b_2 t^3), \qquad\ \  C^{c.t.}= -\ha ( b_1
  s t u + b_2 u^3) \ . \la{238}
\end{equation}
The $b_1$  term  does not contribute  due to the kinematical constraint \rf{211}.\foot{\la{f4} If we relax this constraint 
 and choose $b_1= -\frac{\hat{D}-6}{48 \pi} \Big( \frac{1}{\epsilon} - \gamma + \ln 4 \pi \Big)$ we may cancel 
 the divergent term in \rf{2311} for  any   value of  $stu$, including the  ``evanescent" finite part.
 Let us note also  that  the  $b_1$ term in \rf{237}   may be interpreted as originating from the 
  curvature  integral $\int d^d \sigma \sqrt{-\gamma} R = \int d^d \sigma\  \partial^\beta X^j \partial^\gamma X^k (\partial_\alpha \partial_\beta X^j \partial^\alpha \partial_\gamma X^k - \partial_\beta \partial_\gamma X^j \Box X^k) + \dots$ 
evaluated on the  induced metric in the long-string vacuum \rf{23}. It
% and expanded in derivatives  of the transverse coordinates;
becomes  trivial (integral of a total derivative) in $d=2$. 
 At the same time,  the $b_2$ term  in \rf{237} is the so called Polchinski-Strominger term \ci{Polchinski:1991ax}
 originating from the Polyakov term $\int d^2 \sigma \sqrt{-\gamma}\, R {\nabla}^{-2} R$ 
 evaluated on  \rf{23}  and expanded in derivatives of transverse coordinates
 (see also \ci{Dubovsky:2012sh,Aharony:2013ipa}).
}
Choosing  
\begin{equation}
% b_1= -\frac{\hat{D}-6}{48 \pi} \Big( \frac{1}{\epsilon} - \gamma + \ln 4 \pi \Big), \qquad
 b_2 = -\frac{\hat{D}-24}{96 \pi} \ ,
\end{equation} 
we  then get,  assuming  $t=0, \ u=-s$ as in \rf{235}, 
 \be \la{239}
 \hat A^{(1)}\equiv  A^{(1)}+A^{c.t.}=0 \ , \ \ \ \ 
 \hat C^{(1)} \equiv  C^{(1)}+C^{c.t.}=0\ , \qquad  \hat B^{(1)}\equiv  B^{(1)} +B^{c.t.}=B^{(1)}_f =  \frac{i}{16}  s^3  \ . \ee 
 Let us note 
 that,  in general,  
 the unitarity $\rS^\dagger \rS=1$  of the S-matrix  $\rS = 1 + i \rT $ implies that 
 $2\,  \text{Im} \rT = \rT^\dagger \rT$. In the present case this   leads to the following relations   between 
 the coefficients in \rf{222},\rf{210} and \rf{2133}
\begin{equation} \begin{aligned}
\text{Im}\,  A^{(1)}  & = \frac{1}{4s} \Big[ \hat{D} |A^{(0)}|^2 +  A^{(0)}{} \big(  B^{(0)}{}^* +  C^{(0)}{}^*)  + 
A^{(0)}{}^* \big(  B^{(0)} +  C^{(0)})  \Big]~,\\
\text{Im}\,  B^{(1)} &= \frac{1}{4 s} \Big( |B^{(0)}|^2 + |C^{(0)}|^2 \Big)~, \qquad \qquad 
\text{Im}\,  C^{(1)} = \frac{1}{4 s} \Big( B^{(0)}{}^*| C^{(0)}| + C^{(0)}{}^* |B^{(0)}| \Big)~, \la{246}
\end{aligned}
\end{equation}
which are  indeed satisfied.

\subsection{S-matrix as a pure phase} 
 
 We conclude that for any $\hat D$   the  tree-level  \rf{29}  plus 1-loop \rf{239} 
 scattering amplitude  is given by \rf{222}   where (we restore the dependence on the string tension
 $T_1\equiv T= \tfrac{1}{ 2\pi \a'}$  and assume the $d=2$   kinematics with $t=0$)
\begin{align}\no  
&A= A^{(0)} +  \hat A^{(1)} = 0\ , \qquad \ \ \ \  \qquad C= C^{(0)}  + \hat C^{(1)}=0 \ , \\
%v4-sign
 & B= B^{(0)} +  \hat B^{(1)} = \frac{1}{2\, T} s^2  +\frac{i}{16\, T^2} s^3     \ .  \la{240}
 \end{align}
Thus to 1-loop   level there is no particle creation or annihilation, i.e.  the scattering is purely elastic.

The S-matrix is related to the amplitude $\cal M$ by \rf{a15},\rf{a7}. 
Using  \rf{a8}  in the case  of a 2d  integrable theory with $A=C=0$ 
we conclude  that the S is expressed in terms of  the transmission amplitude $B$ as
\begin{equation}
\rS(\vec{p}_1,\vec{p}_2) = 1+ \frac{i}{4 |\vec p_1 \omega_2 - \vec p_2 \omega_1|}  B[\vec{p}_1,\vec{p}_2,\vec{p}_1,\vec{p}_2]~.\la{242} 
\end{equation} 
For   massless particle scattering   this reduces to
\begin{equation}
\rS = 1 + \frac{i}{2 s} B\ . \la{243} 
\end{equation}
In the present case of \rf{240}  we thus get  at the tree  and 1-loop level 
\begin{equation}
%v4-sign
\rS=  1 +\frac{i}{4\, T} s - \frac{1}{32\, T^2 } s^2   + \OO(T^{-3}s^3) \ . \la{244}
\end{equation}
These are  the first terms in the expansion of the pure-phase (unitary) 
S-matrix 
%v2
 \cite{Dubovsky:2012sh}
 \rf{245}  
 \begin{equation}
 %v4-sign
\rS = e^{ \frac{i }{4T} s }~.\la{245}
\end{equation}

\section{S-matrix on compactified membrane}
\label{s3}

Having  reviewed the computation of the world-sheet S-matrix  in the 
NG string theory, let us perform  a similar   analysis for 
 the  bosonic  membrane  theory.
 
We shall  start  with the Lagrangian \eqref{22}   with  $d=3$.
In the case of the  infinite $\mathbb R^{2} $ membrane in the static gauge 
the  resulting  tree-level  scattering  amplitudes for massless 3d  fields   will  be  again given by \rf{222},
   \rf{210}  where now $s,t$ and $u$   depend on 3d massless  momenta  so that  $s+t+u=0$  but 
   there is no extra 2d condition \rf{211}. 
 
 In 3d  there is no notion of S-matrix  integrability. 
 However,  if we assume  that the membrane   has one compact direction, i.e.
 its   vacuum   configuration  has topology of $\mathbb R \times S^1$, 
 then we may represent the  corresponding 
 3d  world-volume theory as an   effective    string with 2d world sheet
    coupled to an infinite tower of  massive  ``Kaluza-Klein"   2d modes.  We may  then ask if the resulting 2d model   has an  integrable S-matrix once the  contribution of the KK modes is included. 
 
As we shall discuss below,   allowing  massive modes  on the external lines  one does not get an integrable  S-matrix
 already at the tree level. We will also compute the 1-loop S-matrix of scattering of  4 massless  modes 
  generalizing the discussion in the previous section to the case when also an  infinite set of the massive  KK modes 
  are propagating  in the loop.  The resulting  amplitude   will  be  free of  log UV divergences 
   and its  finite part will have a rather  complicated  ``non-integrable" 
      dependence on 2d momenta different from the pure-phase structure \rf{245}
   found  in  the  NG string case.

\subsection{Compactified  membrane  as   effective string  coupled to a massive   tower}
\la{s31}

Let us assume that the classical membrane solution  has topology of $\mathbb R^{1,1} \times S^1$. 
We shall   denote the world-volume membrane coordinates as 
  $(\sigma^0, \sigma^1, \sigma^2)$ and assume that 
 $\sigma^2$ is a  circle of radius $\rR$, i.e.  $\sigma^2 \in [-\pi \rR, +\pi \rR)$.
 We shall   choose a static gauge as (cf. \rf{21},\rf{23})
 \be X^0= \sigma^0, \qquad \ \  X^{D-2}=  \sigma^1, \qquad \ \ 
 X^{D-1}=  \sigma^2\ , \la{3111} \ee
  with $X^j(\sigma^0, \sigma^1, \sigma^2)$, $j=1, ..., D-3$,  being the 
 transverse fluctuation  fields.    Expanding them  in Fourier modes in $\sigma^2$ gives
\begin{equation}
X^j(\sigma^0, \sigma^1, \sigma^2) = % \frac{1}{\sqrt{2 \pi \rR}}
 \sum_{n=-\infty}^{\infty} X^j_n(\sigma^0, \sigma^1) \ e^{\frac{i n}{\rR} \sigma^2}~, \qquad j=1,\dots,  \hat D\equiv D-3~,  \qquad {X}^j_{-n} = (X^j_{n})^*\ , \la{31}
\end{equation}
we thus get an infinite  set of 2d fields $X^j_n(\s^0,\sigma^1) $.
Integrating over $\s^2$ in the membrane action  \rf{21},\rf{22} (using $
\int_{-\pi \rR}^{+ \pi \rR} d \sigma_2 \ e^{i \frac{k}{\rR} \sigma_2} = 2 \pi \rR \delta_{k,0}$)  one gets 
 an effective 2d theory (now $\sigma^\a=(\sigma^0,\sigma^1)$  and 
  $\a, \b=0,1$)\begin{align}
&S = - \hat T \int  d^2 \sigma \Big( 1 + \hat{\mathcal L}_2 + \hat{\mathcal L}_4 + \dots  \Big)~, 
\qquad \qquad    \hat T \equiv   2 \pi \rR  T_2 \ , \la{32}\\
&
\hat{\mathcal L}_2  %  \sum_{n=-\infty}^{\infty}
%\frac{1}{2} (\partial_\alpha X_n^j \partial^\alpha X_{-n}^j + m_n^2 X_n^j X_{-n}^j)
 = \frac{1}{2} \sum_{n=-\infty}^{\infty}   \Big( |\partial_\alpha X_n^j |^2 + m_n^2 |X^j_n|^2 \Big)
 =\ha  (\partial_\alpha X_0^j )^2 + \sum_{n=1}^{\infty}   \Big( |\partial_\alpha X_n^j |^2 + m_n^2 |X^j_n|^2 \Big)
  ~,\ \ \  m_n = \frac{n}{\rR}\ ,\la{33}
\end{align}
Thus there  are $\hat{D}=D-3$  real massless modes $X_0^j$ and an infinite tower of  complex massive modes $X^j_n$. $\hat T$ is the  ``effective  string"  tension. 
The  quartic interaction term $\hat{\mathcal L}_4 $  may be written as (cf. \rf{24}) 
\begin{align}
\hat{\mathcal L}_4 &= %\frac{1}{2 \pi \rR}
 \sum_{n_1,...,n_4=-\infty}^\infty \frac{1}{4}  \Big( c_2 V_{n_1,n_2}^{j,j} V_{n_3,n_4}^{k,k} + c_3 V_{n_1,n_4}^{j,k} V_{n_2,n_3}^{j,k} \Big)\  \delta_{n_1+...+n_4,0}
 ~, \la{34}\\
& V_{n_1,n_2}^{j,k} \equiv 
 \partial_\alpha X_{n_1}^j \partial^\alpha X_{n_2}^k - \frac{n_1 n_2}{\rR^2} X_{n_1}^j X_{n_2}^k~, \la{35}
\end{align}
where $c_2=\ha $ and $c_3=-1$  are the same as in \rf{24}.  
Explicitly,
\begin{equation} \begin{aligned}\la{36}
\hat{\mathcal L}_4 = %\frac{1}{2 \pi \rR}
 \sum^\infty_{n_1,...,n_4=-\infty} \frac{1}{4}  \Big(&c_2 \partial_\alpha X^j_{n_1} \partial^\alpha X^j_{n_2} \partial_\beta X^k_{n_3} \partial^\beta X^k_{n_4} + c_3 \partial_\alpha X^j_{n_1} \partial^\beta X^j_{n_2} \partial_\beta X^k_{n_3} \partial^\alpha X^k_{n_4} \\
&- 2 c_2 \frac{n_3 n_4}{\rR^2}  \partial_\alpha X^j_{n_1} \partial^\alpha X^j_{n_2} X_{n_3}^k X_{n_4}^k -2 c_3 \frac{n_2 n_4}{\rR^2} \partial_\alpha X^j_{n_1} X^j_{n_2} \partial^\alpha X^k_{n_3} X^k_{n_4} \\
&+(c_2+c_3) \frac{n_1 n_2 n_3 n_4}{\rR^4} X^j_{n_1} X^j_{n_2} X^k_{n_3} X^k_{n_4} \Big) \delta_{n_1+...+n_4,0}~.
\end{aligned}
\end{equation}
The massless  $X_0^j$  part here  is   in the first line  and  is % of course
  the same as in the NG action
in \rf{24}.

The necessary  condition for the integrability  of 2d  S-matrix is 
 that there cannot be particle production or annihilation, meaning, in particular, 
  that in any process the number of incoming particles should be the same as the number of outgoing particles. Also, 
  there cannot be any particle transmutation.
   If the  two fields $X_1$ and $X_2$  have different masses then the  processes of the form $X_1+X_1\rightarrow X_2+X_2$ or $X_1+X_2 \rightarrow X_2+X_2$ are forbidden.
    This is because such processes violate macro-causality  (see, e.g., \ci{Dorey:1996gd}). 
    
     Let us now show that for  the interaction 
      Lagrangian \rf{36}  amplitudes for such processes are,  in fact, 
       non-vanishing already at the tree-level.
        Thus  the   theory \rf{32} is not classically integrable.

\subsection{Tree-level 4-point amplitude}
\la{s32}

Let us consider the scattering of four particles (1 and  2 ingoing and  3 and 4 outgoing) with momenta  $p_r$ and mode numbers $n_r$. 
We shall set the compactification radius to one, $\rR=1$,  as  dependence on it is 
easy to restore on dimensional grounds. 
The scattering amplitude  will  be  a generalization of the massless one in   \rf{222}, i.e. 
\begin{equation}\la{37}
\hat{\mathcal M}_{ij,kl} = %\frac{1}{2 \pi} 
\Big(
 \hat{A} \delta_{ij} \delta_{kl} + \hat{B} \delta_{ik} \delta_{jl} + \hat{C} \delta_{il} \delta_{jk}  \Big)\, \delta_{n_1+n_2, n_3+n_4}\ , 
% \delta^{(2)}(p_1+p_2-p_3-p_4) \, \delta_{n_1+n_2, n_3+n_4} \ . %\delta(n_1+n_2,n_3-n_4)~.
\end{equation}
which is multiplied by $\delta^{(2)}(p_1+p_2-p_3-p_4) $ in the S-matrix.
As in the massless  case,  the factor $\hat T^{-1}$ of inverse effective tension in \rf{32}  is implicit here. 
It follows from \rf{36} that there are now five distinct  Feynman diagrams, depending  on the types of particles on external lines  and index contractions (cf. section 2.1)\foot{Here the solid lines represent again propagators with contracted momenta, while dashed lines represent  contractions of target space indices. Since the action \rf{32} including massive modes    contains  also   coupling
  terms  with less  numbers of derivatives,  there are 
   propagators with no contracted momenta. The dotted lines  are used just 
   to indicate the form of the  scattering  diagram. }
%%%%%%%%%%%%
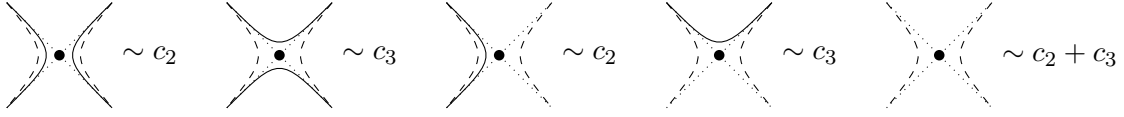
\begin{center}
\begin{tikzpicture}[scale=0.7]
\draw[dotted] (-1,1) -- (1,-1);
\draw[dotted] (-1,-1) -- (1,1);
\draw[-] (-1,1) .. controls (0,0) .. (-1,-1) ; 
\draw[-] (1,1) .. controls (0,0) .. (1,-1) ; 
\draw[dashed] (-1,1) .. controls (-0.2,0) .. (-1,-1) ; 
\draw[dashed] (1,1) .. controls (0.2,0) .. (1,-1) ; 
\fill (0,0)  circle (0.1);
\node[anchor=west] at (1,0) {$\sim c_2$};
\end{tikzpicture}  \quad 
\begin{tikzpicture}[scale=0.7]
\draw[dotted] (-1,1) -- (1,-1);
\draw[dotted] (-1,-1) -- (1,1);
\draw[-] (-1,1) .. controls (0,0) .. (1,1) ; 
\draw[-] (-1,-1) .. controls (0,0) .. (1,-1) ; 
\draw[dashed] (-1,1) .. controls (-0.2,0) .. (-1,-1) ; 
\draw[dashed] (1,1) .. controls (0.2,0) .. (1,-1) ; 
\fill (0,0)  circle (0.1);
\node[anchor=west] at (1,0) {$\sim c_3$};
\end{tikzpicture} \quad
\begin{tikzpicture}[scale=0.7]
\draw[dotted] (-1,1) -- (1,-1);
\draw[dotted] (-1,-1) -- (1,1);
\draw[-] (-1,1) .. controls (0,0) .. (-1,-1) ; 
%\draw[-] (1,1) .. controls (0,0) .. (1,-1) ; 
\draw[dashed] (-1,1) .. controls (-0.2,0) .. (-1,-1) ; 
\draw[dashed] (1,1) .. controls (0.2,0) .. (1,-1) ; 
\fill (0,0)  circle (0.1);
\node[anchor=west] at (1,0) {$\sim c_2$};
\end{tikzpicture}  \quad
\begin{tikzpicture}[scale=0.7]
\draw[dotted] (-1,1) -- (1,-1);
\draw[dotted] (-1,-1) -- (1,1);
\draw[-] (-1,1) .. controls (0,0) .. (1,1) ; 
%\draw[-] (-1,-1) .. controls (0,0) .. (1,-1) ; 
\draw[dashed] (-1,1) .. controls (-0.2,0) .. (-1,-1) ; 
\draw[dashed] (1,1) .. controls (0.2,0) .. (1,-1) ; 
\fill (0,0)  circle (0.1);
\node[anchor=west] at (1,0) {$\sim c_3$};
\end{tikzpicture} \quad
\begin{tikzpicture}[scale=0.7]
\draw[dotted] (-1,1) -- (1,-1);
\draw[dotted] (-1,-1) -- (1,1);
\draw[dashed] (-1,1) .. controls (-0.2,0) .. (-1,-1) ; 
\draw[dashed] (1,1) .. controls (0.2,0) .. (1,-1) ; 
\fill (0,0)  circle (0.1);
\node[anchor=west] at (1,0) {$\sim c_2+ c_3$};
\end{tikzpicture}
\end{center}
We thus get  for the tree-level $A$-amplitude (cf. \rf{210})
\begin{equation} 
\label{3388}
\begin{aligned}
%v4-sign
\hat{A}^{(0)}[p_1,p_2,p_3,p_4] &= -\Big[2 c_2  (p_1 \cdot p_2) (p_3 \cdot p_4) +c_3 (p_1 \cdot p_4)(p_2 \cdot p_3) +c_3 (p_1 \cdot p_3) (p_2 \cdot p_4)  \\
&\quad+ c_3 n_1 n_3 (p_2 \cdot p_4) + c_3 n_1 n_4 (p_2 \cdot p_3) + c_3 n_2 n_3 (p_1 \cdot p_4) + c_3 n_2 n_4 (p_1 \cdot p_3)  \\
&\quad  + 2 c_2 n_1 n_2  (p_3 \cdot p_4)+ 2 c_2 n_3 n_4 (p_1 \cdot p_2) + 2 (c_2+c_3) n_1 n_2 n_3 n_4\Big]~.
\end{aligned}
\end{equation}
Using that $ n_1+n_2=n_3+n_4$  and writing this in terms of 
the Mandelstam variables \eqref{a11}
we get
\begin{equation}\la{39}
%v4-sign
\hat{A}\zz =- \frac{1}{4} (2 c_2 + c_3) \big[s-(n_1+n_2)^2\big]^2 + \frac{1}{2} c_3 \big[t-(n_1-n_3)^2\big]\big[u-(n_1-n_4)^2\big]~.
\end{equation}
The  amplitudes $B$ and $C$   can be obtained using   the crossing relations
\begin{equation}\la{310}
%\hat{A} = \left. 
B= A\Big|_{s \leftrightarrow t, \ n_2 \leftrightarrow -n_3}\ , \ \ \ \ \ \ \ 
C=  A\Big|_{s \leftrightarrow u,\  n_2 \leftrightarrow -n_4}~,
\end{equation}
so that
\begin{equation} \begin{aligned}\la{3200}
%v4-sign
\hat{B}\zz &=- \frac{1}{4} (2 c_2 + c_3) \big[t-(n_1-n_3)^2\big]^2 + \frac{1}{2} c_3 \big[s-(n_1+n_2)^2\big]\big[u-(n_1-n_4)^2\big]~, \\
\hat{C}\zz &= -\frac{1}{4} (2 c_2 + c_3) \big[u-(n_1-n_4)^2\big]^2 + \frac{1}{2} c_3 \big[t-(n_1-n_3)^2\big]\big[s-(n_1+n_2)^2\big]~.
\end{aligned}
\end{equation}
When all the external particles are massless  we recover the result in \rf{210}, 
i.e.   $\hat{A}\zz$ vanishes for the  NG  choice of coefficients $c_2$ and $c_3$ in \eqref{24} and $tu=0$,
\begin{equation}
(\hat{A}\zz)_{0,0\rightarrow0,0} =0.\la{311}
\end{equation} 
 The scattering of  3  massless and 1 massive particle is not allowed by conservation of $n_r$. 
  For 2  massless and 2 massive particles we distinguish two cases. If there is massless + massive  incoming and  massless + massive outgoing  states 
 then the  kinematic constraint is $t=0$ for $n_3=n_1=n$, $n_2=n_4=0$ or $n_4=n_2=n$, $n_1=n_3=0$, which leads to
\begin{equation}\la{312}
(\hat{A}\zz)_{n,0\rightarrow n,0} = (\hat{A}\zz)_{0,n\rightarrow 0,n} = 0.
\end{equation}
The following amplitudes are, however, non-zero:
\begin{equation}\la{314}
%v4-sign
(\hat{A}\zz)_{n,0 \rightarrow 0,n} = (\hat{A}\zz)_{0,n \rightarrow n,0} =- \frac{n^2(s-n^2)^3}{2s^2}.
\end{equation}
The second case is when both incoming (or both outgoing) 
particles are massless. Then  $n_1=n_2=0$  and $n_3=-n_4=n$   and 
 the relation between
 the Mandelstam variables  in  
 \rf{a12}   implies that  $t u = n^4$. 
Consequently, for the NG  values of   $c_2$ and $c_3$ in \rf{24}
\begin{equation}\la{315}
%v4-sign
(\hat{A}\zz)_{0,0 \rightarrow n, -n}  =- \frac{n^2}{2} s.
\end{equation}
The non-zero value   of this amplitude 
 indicates that the theory is not integrable:  this is  an example of particle transmutation  when  2 massless particles scatter
  into 2  massive ones. 
The non-zero result for   $\hat{A}\zz$  is found  (using again \rf{a12})  also for 
amplitudes  with 3  and 4  massive particles. 
Thus the resulting  tree-level S-matrix is not integrable  for all  processes  
with massive particles on  external lines.

\subsection{1-loop contribution to the  massless 4-point amplitude}\la{s33}

Let us now compute the 1-loop   contribution to the scattering  of 4 massless ($n_r=0$) 
modes.
Similarly to the string case in section 2.2  this 
  amplitude  is   given by  the   contributions of 
 the  3  bubble diagrams  in Figure \ref{fig:bubble}  but   now  all  (massless  and massive)  modes  appear in internal lines.
 The amplitude takes the same form as  in \rf{2133}
\begin{equation}\la{316}
\hat{\mathcal M}^{(1)}_{ij,kl} = %  \Big(
\hat{A}^{(1)} \delta_{ij} \delta_{kl} + \hat{B}^{(1)} \delta_{ik} \delta_{jl} + \hat{C}^{(1)} \delta_{il} \delta_{jk}
%\Big)\,  \delta^{(2)}(p_1+p_2-p_3-p_4)
 ~, 
\end{equation}
entering the S-matrix  together with a $ \delta^{(2)}(p_1+p_2-p_3-p_4)$  factor. 

From the properties of the 4-vertex, both legs in the loop have opposite mode numbers (and therefore the same mass). 
 The total   $s$-channel  amplitude  is given  by the sum 
 of contributions of  modes with fixed mode number propagating in the loop, 
 \begin{equation}
\hat{A}_s^{(1)} = \sum_{n=-\infty}^{\infty} \hat{A}_{n,s}^{(1)} \ , \la{319}
\end{equation}
where the  mode $n$ 
 amplitude   has   the form (cf. \rf{215})   %(here written for the $s$-channel)
\begin{align}
\hat{A}^{(1)}_{n,s} &= \frac{1}{2i} \int \frac{d^d p}{(2 \pi)^d} \frac{\hat N_s}{(p^2 + n^2)[(p_1+p_2-p)^2 + n^2]}  =  \frac{1}{2i} \int_0^1 dx \int \frac{d^d  l}{{(2 \pi)^d} } \frac{\hat N_s}{(l^2 + \hat \Delta_s)^2}~, \la{317}
\\
& l=p-(p_1+p_2) x, \qquad  \hat \Delta_s = n^2  + \Delta_s\ , \ \ \ \ \ \ 
\Delta_s=  - x(1-x) s~. \la{318}
\end{align}
Here   we again use dimensional regularization with   $d=2-2\epsilon$ and
 implicitly  assume  the presence  of the  $-i\ve$ shift as  in \rf{215},\rf{512} 
and also  the overall  tension (coupling) factor  of $ \hat T^{-2}$ from \rf{32}. 
As in the tree-level  amplitude  \rf{3388},\rf{39}
  we also set $\rR=1$ in \rf{33},\rf{36}.   
  As  the $S^1$ radius  $\rR$ enters  the Lagrangian \rf{32}--\rf{36}  via $\frac{n_i}{ \rR}$, 
  it can be restored  by rescaling the mode number 
    \be \la{555} n\  \to\  \frac{n}{\rR}
   \ ,  \ee 
 or equivalently by  rescaling $(s,t,u) \rightarrow \R^2 (s,t,u)$ (implying in particular $\Delta_s \to \rR^2 \Delta_s$) and adding an overall factor in the amplitude. 
   
The numerator $\hat N_s$ in \rf{317} 
 has the same form  as in \eqref{219}, but the tree-level amplitudes $A\zz, B\zz, C\zz$ there should be now  replaced with their ``massive'' counterparts $\hat{A}\zz,\hat{B}\zz,\hat{C}\zz$, as given in \eqref{3388}. 
 The expansion of  $\hat N_s$ in powers of $l$  has the same form as in \rf{288}
 where now (cf. \rf{2199})
 \begin{align}
&N_0 = -s^3 x(1-x) n^2 + \frac{1}{4} (\hat{D}-4) s^2 n^4~,  \quad 
N_2 = -s^3 x(1-x)+ \frac{1}{2} (\hat{D}-4) s^2 n^2~,  \quad 
N_4 = \frac{1}{4}(\hat{D}-4)  s^2~, \no \\
&N_{\mu \nu} = \big[ -2 s^2 x(1-x) + s(\hat{D}-2) n^2 \big] \big(p_{1,\mu} p_{2,\nu}+ p_{3,\mu}p_{4,\nu} \big)~,\la{320} \\
&M_{\mu \nu} = (\hat{D}-2) s \big(  p_{1,\mu} p_{2,\nu}+ p_{3,\mu}p_{4,\nu}  \big) ~, \qquad 
N_{\mu \nu \rho \sigma} = 4 \hat{D}\, p_{1,\mu} p_{2,\nu} p_{3,\rho}p_{4,\sigma }  \ . 
\no \end{align}
In  the $t$-channel (and similarly in the $u$-channel with $t \rightarrow u$) 
 we find that
\begin{align}
& N_0 = \frac{1}{2} t^2 \big[ t x(1-x)+n^2 \big]^2~, \quad N_2 = t^3 x(1-x) + t^2 n^2~, \qquad N_4 = \frac{t^2}{2}~, \quad 
 M_{\mu \nu} = 0~, \quad N_{\mu \nu \rho \sigma} = 0~, \no \\
& \begin{aligned} N_{\mu \nu} &=  -2 t^2 x (1-x) \Big(
p_{1,\mu} p_{4,\nu} - p_{1,\mu} p_{2,\nu} - p_{3,\mu} p_{4,\nu}  + p_{3,\mu} p_{2,\nu}
\big)  - t^2 \big(p_{1,\mu} p_{2,\nu}  + p_{3,\mu} p_{4,\nu}    \big)~.
\end{aligned}
\end{align}
From these expressions  we can then compute the coefficients $\gamma_k$
defined as in \rf{223}  and 
appearing in the analogs of the expressions \rf{221} and \rf{224}.
Remarkably, we find that they do not depend   not only  on 
$d=2-2\eps $  but also  on the mode number $n$
and we find  the expressions that look the same as in the massless  case in \rf{2255}, i.e.  %in the $s,t$ and $u$ channels
\begin{align} %\label{eq:gamma_s}
\GG_{0,s}& = \frac{1}{4} (\hat{D}-8) 
s^4 x^2 (1-x)^2~, &\quad \GG_{2,s} &= -s^3 x(1-x)~, &\quad \GG_{4,s} &= s^2 - \frac{1}{2} \hat{D} t u~,\la{3223}
\\
\GG_{0,t} &= 2 t^4 x^2 (1-x)^2~, &\quad  \GG_{2,t} &= \frac{1}{2} t^2 s + 3 x(1-x) t^3~, &\quad \GG_{4,t} &=0~, \la{32231}\\
\GG_{0,u} &= 2 u^4 x^2 (1-x)^2~, &\quad  \GG_{2,u} &= \frac{1}{2} u^2 s + 3 x(1-x) u^3~, &\quad \GG_{4,u}&=0~.\la{32232}
\end{align}
Some details of 
 the calculation of the  1-loop amplitudes at fixed  $n$ 
 can be found in Appendix \ref{a4}.

Expanding the analog of \rf{221} in $\eps\to 0$  and summing over the three channels  as in \rf{213} 
we get  for the singular part of 
the   $\hat{A}_{n}^{(1)}$ amplitude: 
\begin{equation} \label{325}
\hat{A}_{n,\epsilon}^{(1)} = \frac{1}{96 \pi} \Big( \frac{1}{\epsilon} - \gamma + \log 4 \pi \Big) \Big[ - (\hat{D}-6) stu -12 n^2 \big(s^2 - \tfrac{1}{2} \hat{D} tu \big) \Big]~.
\end{equation}
As we have only massless  particles on external legs, the $stu$ term  vanishes due to the kinematical  constraint \rf{211}  and the divergence is thus 
proportional to  $n^2$ or the effective mass-squared   term. 

To sum over $n$ we   will  apply   the   Riemann $\zeta$-function regularization. 
 The same   regularization was used 
  in the 1-loop membrane  computations in \ci{Giombi:2023vzu,Beccaria:2023ujc}.  It  is consistent  with   the
  expected  absence of the  1-loop logarithmic divergences in the original 3d  membrane  theory 
  (the absence of 1-loop  log UV   divergences  in a 3d theory 
  should not  depend on $S^1$ compactification). Explicitly,   for any positive  or zero integer 
  $k$  we will set\foot{The existence of these ``trivial"  %$z=-2k$ 
zeroes  of $\zeta_R(z) =\sum_{n=1}^\infty n^{-z}$ follows, e.g.,  from the reflection formula 
$\zeta_R(z) = 2^z \pi^{z-1}  \sin \frac{\pi z}{ 2}\, \Gamma( 1-z)\,   \zeta_R(1-z)$ 
after setting $z=-2k$. 
This   property can be proved also directly    by  considering the regularized sum 
$\sum^\infty_{n=1}  n^{2k} \, e^{-\ve  n} = \frac{\del^{2k}}{ \del \ve^{2k}} I(\ve) $
where $I(\ve) = \sum^\infty_{n=1}  e^{-\ve  n}= \frac{1}{ 1 -  e^{-\ve  n}} $ 
 and observing  that the resulting expansion in $\ve\to 0$   contains only odd powers of $\ve$. Indeed, 
 $ \frac{\del^{2k}}{ \del \ve^{2k}} I(\ve)$  changes sign under $\ve\to -\ve$ 
 as $ \frac{1}{ 1 -  e^{-\ve  n}}  = - \frac{1}{ 1 -  e^{\ve  n}} +  1 $    so  that 
 and thus  there is  no finite part  after subtracting all   poles.
}
\begin{align}
\sum_{n=-\infty}^\infty n^{2k}= 2\,  \zeta_R(-2k) =0 \ , \la{326}\qquad \qquad 
\sum_{n=-\infty}^\infty1=
  1 + 2  \zeta_R(0) =0 \ . \end{align}
   Thus  the sum of the 
   full expression \rf{325}  is regularized  to 0 even  without using that  $stu=0$. 
 %  the 1-loop amplitude is finite. 

  Thus   the total 1-loop 
   amplitude  is finite  and given   by the sum over $n$ of the remaining parts of the   partial 
   amplitudes (we again choose the kinematics so that $t=0$ and thus $u=-s$)
\begin{align}
\hat{A}^{(1)}_{n} &= - \frac{s^2}{192 \pi} \Big[(\hat{D}-24) s  + 
6    (\hat{D}+4) n^2  - 24 n^2  \ln n^2 
- 6  n^2     (\hat{D} n^2 - 2 s) Q_n(-s)  + 12 n^2 s Q_n(s) \Big]~, \no \\
\hat{B}^{(1)}_{n} &= \frac{s^2}{192 \pi} \Big[ - 12 \hat{D} n^2  + 12 \hat D n^2  \ln n^2 
  +   6 s(s-2 n^2) Q_n(-s) + 6 s (s+2 n^2) Q_n(s)  \Big]~, \la{330}\\
\hat{C}^{(1)}_{n} &= \frac{s^2}{192 \pi} \Big[(\hat{D}-24) s - 6 
  (\hat{D}+4) n^2   + 24 n^2  \ln n^2 
  + 6 n^2 (\hat{D} n^2 +2 s)  Q_n(s)  +12 n^2  s Q_n(-s)  \Big]~.\no 
\end{align}
 Here the function  $Q_n(s)$ is defined by (see \rf{d8})
\begin{gather}\la{3322}
Q_n(s) \equiv  - \frac{2}{s \sqrt{1+\frac{4 n^2}{s}}} \ln   \frac{ \sqrt{1+\frac{4 n^2}{s}}-1}{\sqrt{1+\frac{4 n^2}{s}}+1}\ , \\
Q_n(s)\Big|_{n\to \infty} = \frac{1}{n^2} -\frac{s}{6 n^4} + \OO(\frac{1}{ n^6}) \ , \qquad Q_n (s)\Big|_{n\to 0}  = -\frac{2}{s}\, \ln \frac{n^2}{s}  +\OO(n^2) 
\ . \la{2317}
\end{gather}
One can check that for $n=0$  the expressions in \rf{330} 
 reduce to the ones  for the NG string   in \rf{232}--\rf{234}
  (with $t=0$).
  
  Note that restoring  the $\rR$ dependence in \rf{330} using \rf{555}  as well  as the effective tension factor  from \rf{32} 
   corresponds to 
   \be \la{5555}
   %s\to \rR^2 s \ , \qquad \qquad   
   \hat{A}^{(1)}_{n} (s)\  \ \to \ \   \hat T^{-2}\, \rR^{-6}  \hat{A}^{(1)}_{n} ( \rR^2 s) \ , \qquad  \ {\rm etc} 
   \ . \ee

  \subsubsection{Finite expression for the amplitude}\la{s331}
  
   The  total   amplitude  is thus given by 
 \be \la{334}
 \hat{A}^{(1)}= \hat{A}^{(1)}_0  + 2 \sum_{n=1}^\infty \hat{A}^{(1)}_{n,f} \ ,  \ee
 and by  similar expressions for $\hat{B}^{(1)}$ and $\hat{C}^{(1)}$.
 We shall   define the sum  over $n$  using again the Riemann  $\zeta$-function. Then 
 using \rf{326} the terms  with only  $n^{2k}$ factors  in \rf{330} 
  will   be  zero  after the summation.  We  will also need  % use also   %may use that 
 \be \la{335}
 %\sum_{n=1}^\infty   \log n \ \to \  - \zeta'_R (0) = \ha \log (2 \pi) \ , \qquad  \qquad \ \  \ 
 \sum_{n=1}^\infty   n^2 \log n  =  - \zeta'_R (-2) = 0.0305...  \ . \ee
  The  remaining non-trivial sums   involving  $Q_n(s)$
  can be defined  by extracting the  part that diverges  at large $n$ and define
  it using \rf{326},\rf{335}.  As  the expressions in \rf{330}   contain  $Q_n$   multiplied 
   by $n^{2k}$ with  $k=0,1,2$,  then  in view of % the large $n$ expansion in 
   \rf{2317}   it is useful to  subtract 
   the $1\ov n^2$ and $1\ov n^4$ parts from the  large $n$  expansion  of $Q_n$  and  define 
   (for $n >0$) 
   \be 
   \la{336}
   {\bar Q}_n(s)=  Q_n(s) -  \frac{1}{n^2} + \frac{s}{6 n^4}   %, \qquad \qquad n >0 
  \ . \ee 
  Then   the  sums involving $\bar Q_n$  will be finite, i.e. the non-trivial part of the amplitude will   be given by finite  terms like 
  $\sum_{n=1}^\infty ( an^4 + b n^2 + c) {\bar Q}_n(s)$.  Explicitly,   written in 
 terms of  $\bar Q_n$   the expressions in \rf{330} read 
 \begin{align}
 \hat{A}^{(1)}_{n,f} &= - \frac{s^2}{192 \pi} \left[24 n^2 - 24 n^2 \ln n^2 - 6 \hat{D} n^4 \bar{Q}_n(-s) + 12 n^2 s \big(\bar{Q}_n(-s) + \bar{Q}_n(s)\big) \right]~, \no \\
 \hat{B}^{(1)}_{n,f} &= \frac{s^2}{192 \pi} \left[-12 \hat{D} n^2 + 12 \hat{D} n^2 \ln n^2 + \frac{8 s^2}{n^2} + 6 s (s-2 n^2) \bar{Q}_n(-s) + 6s (s+2 n^2) \bar{Q}_n(s)\right]~, \la{3344}\\
 \hat{C}^{(1)}_{n,f} &= \frac{s^2}{192 \pi} \left[-24 n^2 +24 n^2 \ln n^2 + 6 \hat{D} n^4 \bar{Q}_n(s) + 12 n^2 s \big(\bar{Q}_n(-s) + \bar{Q}_s(s)\big) \right]~.\no 
 \end{align}
As a result, we find (here $\zeta_R (2) = {\pi^2\ov 6}$) 
 \begin{align}
 \hat{A}^{(1)} &= - \frac{\hat{D}-24}{192 \pi} s^3 - \frac{s^2}{16 \pi} \Big[
 8 \, \zeta'_R(-2)  - \hat{D}\, \P_4 (-s)  + 2 s  \big( \P_2(s) + \P_2(-s)\big )  \Big]~, \la{3355a}  \\
 \hat{B}^{(1)} &=   \frac{i}{16} s^3 +
  \frac{s^2}{16 \pi} \Big[- 4 \hat{D}\, \zeta_R'(-2) + \frac{4}{3} s^2 \zeta_R(2) \no \\
  & \qquad \qquad \qquad \qquad + 
    s^2  \big( \P_0(s) + \P_0(-s)\big ) 
   + 2 s \big( \P_2(s) -\P_2(-s)\big )
  %2\sum_{n=1}^\infty \left( 6 s (s-2 n^2) \bar{Q}_n(-s) + 6s (s+2 n^2) \bar{Q}_n(s)\right)
  \Big]~,\la{3355b} \\
 \hat{C}^{(1)} &= \frac{\hat{D}-24}{192 \pi} s^3 + \frac{s^2}{16 \pi} \Big[
  -8 \, \zeta'_R(-2)  + \hat{D}\, \P_4(s)   + 2 s  \big( \P_2(s) + \P_2(-s)\big ) \Big] \ , \la{3355c}\\
  &\qquad \qquad \qquad   \P_{2k} (s) \equiv  \sum_{n=1}^\infty n^{2k}  \bar{Q}_n(s) \ . \la{3366}
  % \sum_{n=1}^\infty n^4 \bar{Q}_n(s) + 24 s \sum_{n=1}^\infty n^2 (\bar{Q}_n(-s)+\bar{Q}_n(s)) \Big]~.
 \end{align}
    We conclude that  the 1-loop 4-point massless  scattering amplitudes in  compactified   membrane theory    are 
   expressed in terms of   complicated    (non-polynomial) functions of $s$
   that 
     cannot be cancelled   by  adding  local counterterms. This 
    is again  an indication of non-integrability of the compactified membrane  theory
    viewed as an effective 2d theory.

%   \subsubsection{An alternative representation}\la{s332}
   
   One may   also   obtain  an alternative representation for the 1-loop   amplitude 
   by   starting  with the Feynman  parameter 
    integral representation  \rf{317}   and first doing the   sum  over $n$
    under the integral. This gives 
\begin{equation}\la{338}
\hat{A}^{(1)}_s = \frac{1}{8 \pi} \int_0^1 dx \Big[ \GG_{0,s} F(1+\epsilon; \Delta_s) + \GG_{2,s} F(\epsilon; \Delta_s) + \GG_{4,s} F(-1+\epsilon; \Delta_s)  \Big]~, 
\end{equation}
where  $\Delta_s= - x (1-x) s $  and $\GG_{k,s}$   were given in \rf{318}  and \rf{3223},    and 
 we defined a function $F$  which is   proportional to the Epstein zeta function
\begin{equation}
F(w;c )\equiv  \Gamma(w)\,  \zeta_E(w; c )~,\qquad  \qquad  \zeta_E(w; c) = \sum_{n=-\infty}^{\infty} \frac{1}{(n^2 +c )^w}~.\la{339} 
\end{equation}
To understand the pole structure of the function $F(w;c)$  in $w$ 
 it is useful to use 
  its infinite sum representation in  \rf{b7} (see  \rf{b6},\rf{b7})
  \be \la{bb7}
  F(w;c)
 ={ \sqrt{\pi} \ov c^{w-\ha} } \Gamma(w-\tfrac{1}{2} )+ {4 \pi^w\ov  (\sqrt{c})^{w- \ha}} 
 \sum_{n=1}^\infty l^{w-\ha } K_{w- \ha }(2 \pi l \sqrt{c})\ , 
\end{equation}
where  we assumed that $c>0$   and 
%\begin{equation} \label{eq:Fsum}
%F(w;\Delta)= \sqrt{\pi} \Gamma(w-1/2) \Delta^{1/2-w} + 4 \pi^w \sqrt{\Delta}^{1/2-w} \sum_{l=1}^\infty l^{w-1/2} K_{w-1/2}(2 \pi l \sqrt{\Delta})~,
%\end{equation}
 $K_\nu$ denotes the modified Bessel function of the second kind.\foot{The first term in \eqref{bb7} contains poles at $w={1\ov 2}, -{1\ov 2},-{3\ov 2}, \dots$.
 The second term is a convergent series.}  This  expression provides an  analytic continuation of the Epstein zeta function and can be used to regularise the infinite sum for $w \leq 0$ 
  (in \rf{338} we need $w\to 0$ and $w\to -1$).
   Using \rf{bb7} 
   we can then take the limit 
    $\epsilon \rightarrow 0$    and  end up with the following  finite expression\foot{
    It should be possible  to relate this representation to the one in \rf{3355a}-\rf{3355c}. Note, in particular, that 
    the sums $\P_{2k}(s)$  should   admit  equivalent  integral representations
 similar to the following  one for the   sum of $Q_n$ 
 \begin{equation}\no
% \P_0 (s) = ....  {\rm give   this instead  of } 
 \sum_{n=-\infty}^{\infty} Q_n(s) = \sum_{k=0}^\infty \left( \frac{s}{4} \right)^{k} \frac{1}{2k+1} \zeta_E\Big(k+1;\frac{s}{4}\Big) = \int_0^1 d x \, \zeta_E\big(1; x (1-x)s \big)\ , 
 \end{equation}}
 \begin{align}\la{341}
\hat{A}^{(1)}_s &=  \frac{1}{8 \pi} \int_0^1 dx   \Big[\GG_{0,s} F(1;\Delta_s) +\GG_{2,s} F(0;\Delta_s) + \GG_{4,s} F(-1;\Delta_s)  \Big]~, 
\\
F(1;\Delta_s) &= \frac{\pi}{\sqrt{\Delta_s}} + \frac{2 \pi}{\sqrt{\Delta_s}} \,\text{Li}_0(e^{-2 \pi \sqrt{\Delta_s}})~, \qquad 
F(0;\Delta_s) = -2 \pi \sqrt{\Delta_s}+ 2 \,\text{Li}_1(e^{-2 \pi \sqrt{\Delta_s}})  ~,\la{342} \\
F(-1;\Delta_s) &= \frac{4}{3} \pi (\sqrt {\Delta_s})^3 %^{3/2} 
+ \frac{2 \sqrt{\Delta_s}}{\pi} \,\text{Li}_2(e^{-2 \pi \sqrt{\Delta_s}}) + \frac{1}{\pi^2} \text{Li}_3 (e^{-2 \pi \sqrt{\Delta_s}})~.\la{343}
\end{align}
Let us recall that here $\Delta_s = - x (1-x) s$   with 
$x \in [0,1]$ and  $s>0$, so that $\Delta_s <0 $  in the physical kinematic region. 
To define $\sqrt{\Delta_s}$  let us 
 recall that we implicitly assume  the $i \varepsilon$  shift  in the propagators in \rf{317} 
 as in \rf{215},\rf{512},  i.e. 
\begin{equation} \label{eq:deltae}
\sqrt{\Delta_s} \rightarrow \sqrt{\Delta_s-i \varepsilon} = \sqrt{-x(1-x) s - i \varepsilon} = -i \sqrt{|\Delta_s|} + \tilde{\varepsilon}\ ,\ \  \qquad \tilde{\varepsilon}>0~.
\end{equation} 
Similar   finite expressions  can be  found  for $\hat{A}^{(1)}_t $   and $\hat{A}^{(1)}_u$ 
using  the values of $\gamma$-coefficients in 
\rf{32231}  and \rf{32232}.

\subsubsection{Tadpole  diagram  contribution} 

In addition to  the  bubble diagram contributions (cf. Figure \ref{fig:bubble})
 leading to \rf{317}  there   is  also  a possible   1-loop  tadpole
  contribution. It  was vanishing in the string case in section 2.2
  but may be non-vanishing in the present 
   case of the massive modes propagating in the loop. 

There are five different possibilities for the contraction of indices in the 6-point 
vertex in the action in \rf{27},\rf{32} 
 that  contribute to the one-loop 4-point amplitude $A^{(1)}$
 (where  particles 1, 2   and 3, 4 have contracted indices).
 Pictorially, these are given by 
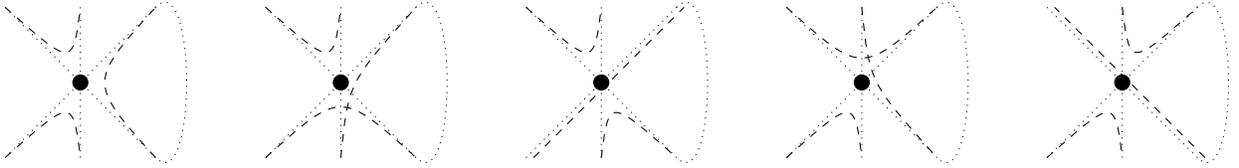
\begin{center}
\begin{tikzpicture}
\draw[dotted] (-1,-1)  -- (0,0);
\draw[dotted] (0,-1) -- (0,0);
\draw[dotted] (1,-1)-- (0,0);
\draw[dotted] (-1,1) -- (0,0);
\draw[dotted] (0,1)  -- (0,0);
\draw[dotted] (1,1)  -- (0,0);
\draw[dotted] (1,1) to [out=45,in=-45]   (1,-1);
\draw[fill] (0,0) circle (0.1);

\draw[dashed] (-1,-1) .. controls (-0.1,-0.2) .. (0,-1); 
\draw[dashed] (1,-1) .. controls (0.1,0) .. (1,1); 
\draw[dashed] (-1,1) .. controls (-0.1,0.2) .. (0,1); 

%\node[] at (0,-1.5) {$A[p_1,p_2,p,p,p_3,p_4]$};
\end{tikzpicture} \qquad
\begin{tikzpicture}
\draw[dotted] (-1,-1)  -- (0,0);
\draw[dotted] (0,-1) -- (0,0);
\draw[dotted] (1,-1)-- (0,0);
\draw[dotted] (-1,1) -- (0,0);
\draw[dotted] (0,1)  -- (0,0);
\draw[dotted] (1,1)  -- (0,0);
\draw[dotted] (1,1) to [out=45,in=-45]   (1,-1);
\draw[fill] (0,0) circle (0.1);

\draw[dashed] (-1,-1) .. controls (0,-0.1) .. (1,-1); 
\draw[dashed] (0,-1) .. controls (0.1,0) .. (1,1); 
\draw[dashed] (-1,1) .. controls (-0.1,0.2) .. (0,1); 

%\node[] at (0,-1.5) {$A[p_1,p,p_2,p,p_3,p_4]$};
\end{tikzpicture}  \qquad
\begin{tikzpicture}
\draw[dotted] (-1,-1)  -- (0,0);
\draw[dotted] (0,-1) -- (0,0);
\draw[dotted] (1,-1)-- (0,0);
\draw[dotted] (-1,1) -- (0,0);
\draw[dotted] (0,1)  -- (0,0);
\draw[dotted] (1,1)  -- (0,0);
\draw[dotted] (1,1) to [out=45,in=-45]   (1,-1);
\draw[fill] (0,0) circle (0.1);

\draw[dashed] (-1+0.1,-1) .. controls (0.1,0) .. (1+0.1,1); 
\draw[dashed] (0,-1) .. controls (0.1,-0.2) .. (1,-1); 
\draw[dashed] (-1,1) .. controls (-0.1,0.2) .. (0,1); 

%\node[] at (0,-1.5) {$A[p_1,-p,p,-p_2,p_3,p_4]$};
\end{tikzpicture} \qquad
\begin{tikzpicture}
\draw[dotted] (-1,-1)  -- (0,0);
\draw[dotted] (0,-1) -- (0,0);
\draw[dotted] (1,-1)-- (0,0);
\draw[dotted] (-1,1) -- (0,0);
\draw[dotted] (0,1)  -- (0,0);
\draw[dotted] (1,1)  -- (0,0);
\draw[dotted] (1,1) to [out=45,in=-45]   (1,-1);
\draw[fill] (0,0) circle (0.1);

\draw[dashed] (-1,-1) .. controls (-0.1,-0.2) .. (0,-1); 
\draw[dashed] (-1,1) .. controls (0,0.1) .. (1,1); 
\draw[dashed] (0,1) .. controls (0.1,0) .. (1,-1); 

%\node[] at (0,-1.5) {$A[p_1,-p,p,-p_2,p_3,p_4]$};
\end{tikzpicture} \qquad
\begin{tikzpicture}
\draw[dotted] (-1,-1)  -- (0,0);
\draw[dotted] (0,-1) -- (0,0);
\draw[dotted] (1,-1)-- (0,0);
\draw[dotted] (-1,1) -- (0,0);
\draw[dotted] (0,1)  -- (0,0);
\draw[dotted] (1,1)  -- (0,0);
\draw[dotted] (1,1) to [out=45,in=-45]   (1,-1);
\draw[fill] (0,0) circle (0.1);

\draw[dashed] (1+0.1,-1) .. controls (0.1,0) .. (-1+0.1,1); 
\draw[dashed] (-1,-1) .. controls (-0.1,-0.2) .. (0,-1);
\draw[dashed] (0,1) .. controls (0.1,0.2) .. (1,1); 

%\node[] at (0,-1.5) {$A[p_1,-p,p,-p_2,p_3,p_4]$};
\end{tikzpicture} 
\end{center}
The first diagram contains an internal loop over the $SO(\hat{D})$ index and therefore contributes a factor of $\hat{D}$. 
As a result, the tadpole  contribution to the fixed $n$  amplitude $A^{(1)}_{{\rm tad}, n}$ 
may be written as 
\begin{gather}
 A^{(1)}_{{\rm tad}, n}= -\frac{1}{2 i} \int \frac{d^d p}{(2 \pi)^d} \frac{N_{\rm tad} }{p^2+n^2-i \varepsilon}\ , \la{3442}
\\ 
\begin{aligned}
& N_{\rm tad} = \hat{D} A\zz[p_1,p_2,p,p,p_3,p_4]  + A\zz[p_1,p,p_2,p,p_3,p_4] \\
& \ \ \ \ \ \ \ \ \  + A\zz[p_1,-p,p,-p_2,p_3,p_4] +A\zz[p_1,p_2,p,p_3,p,p_4] + A\zz[p_1,p_2,p,p_4,p_3,p]\ ,
\end{aligned}  \la{3443} 
\end{gather}
%\end{equation}
where $A\zz[p_1,p_2,p_3,p_4,p_5,p_6]$   is  the tree-level 6-point amplitude  in \rf{eq:Atree6}. 
Setting   $d=2-2 \epsilon$  and taking $\eps\to 0$ this leads to
%%%%%%%
\iffa 
\begin{equation}
A^{(1)}= -\frac{1}{4 \pi} \Big[ \big( \frac{1}{\epsilon}- \gamma + \ln 4 \pi \big)  -  \ln n^2  \Big] n^2 (2 (2 c_4 + c_5 + c_6)s^2 -\frac{2}{3} (4 c_5+c_6) t u  +\frac{\hat{D}}{6}(4 c_5 + 3 c_6) s^2 - \hat{D} c_6 t u )
\end{equation}
with $n$ the mode number of the particle in the internal loop.
\fi
%%%%%%%%%
\begin{equation}
A^{(1)}_{{\rm tad}, n}= \frac{1}{8 \pi} \Big[ \big(\frac{1}{\epsilon}- \gamma + \ln 4 \pi \big)  -  \ln n^2 + 1  \Big] \,  n^2 (s^2 - \hat{D}\,  t u )\ . \la{3466}
\end{equation}
Summing over $n$ using \rf{326},\rf{325}  we get 
 a finite term \foot{Being polynomial in momenta, \rf{6643} 
  can be, in principle, cancelled by a local counterterm.}      
 \be\la{6643}
 A^{(1)} _{\rm tad} = \frac{1}{2\pi} \zeta_R'(-2) \, (s^2 - \hat{D}\, t u ) \ . 
 \ee 
  This   contribution  should be  combined with the bubble diagram  one in \rf{3355a} (where $t=0$);  
   as a result  the total amplitude does not contain the $\zeta_R'(-2) s^2$ term. 
   The same applies to $B^{(1)}$ and $ C^{(1)} $ amplitudes.

\subsubsection{\texorpdfstring{$\rR \to \infty$}{R->infinity} and \texorpdfstring{$\rR \to 0$}{R->0}  limits}

Let   us now   restore the dependence on the   compactification radius $\rR$ 
using \rf{555}
and consider the  limits $\rR \to \infty$ and $\rR \to 0$, or, equivalently,  large and small   momenta. For large $\rR$ or large momenta  we should expect to recover 
the  scattering  amplitude  on 
 uncompactified (i.e. $\RR^2$-shape)   membrane, while 
for small $\rR$ or small momenta the  contributions of  massive KK modes  should be suppressed  and we should  get the scattering amplitude of  the NG  string.

%\begin{equation}
%\hat{A}_s^{(1)} \rightarrow \rR^{-6} \hat{A}_s^{(1)}, \qquad (s,t,u) \rightarrow \rR^2 (s,t,u).
%\end{equation}

As follows from  the form of the action in \rf{32},\rf{33},\rf{36}   for $\rR\to \infty$ 
we need also  to  consider  $n$ large   and introduce  as usual a continuous 3rd component of the momentum 
$p_3 = {n\ov \rR}$  thus  effectively recovering the  case  of uncompactified membrane.
To take this  limit directly in the amplitude \rf{341}--\rf{343}  
with   $\Delta _s \to \rR^2 \Delta_s$  %the momenta  rescaled as in \rf{555}  
we note that 
using \eqref{eq:deltae} we   have  (keeping $\tilde{\varepsilon} >0$)
 \begin{equation}
\lim_{\rR \rightarrow \infty} \text{Li}_k (e^{-2 \pi \rR \sqrt{\Delta_s}}) =  \lim_{\rR \rightarrow \infty} \text{Li}_k (e^{2 \pi i \sqrt{|\Delta_s|}- 2 \pi \tilde{\varepsilon} \rR}) =0~, \qquad k=-1,0,1~\ . \la{348}
 \end{equation}
 %where we kept $\tilde{\varepsilon} >0$ small but non-zero. 
 Thus in the large $\rR$ limit 
the terms involving polylogarithms  in \rf{342},\rf{343}  do not contribute to the integral   over the Feynman parameter  in \rf{341}.  

 Including the tension factor 
  $\hat T^{-2}= (2\pi \rR)^{-2}  T^{-2}_2 $  from \rf{32} as in \rf{5555}  
  (with  $T_2$  held  fixed in the large $\rR$  limit) and  keeping 
 $s,t$   arbitrary   (with $s+t+u=0$) we thus  get  for the limit of the  total 
 $\hat{A}^{(1)}$  amplitude
 %\foot{\la{f77}Note that   as $ T_2\sim$ (length)$^{-3}$ 
 %this amplitude has   dimension  (length)$^{-1}$  as appropriate for  $d=3$
 %(cf. \rf{a15},\rf{a7}).}
\be
%v3
   \hat{A}^{(1)}\Big|_{\rR\to \infty}   = \frac{1}{2 \pi \rR} {1\ov 256\, T^2_2}  
  \Big[ (-s)^{3/2}  \big[
 (\tfrac{3}{32}  \hat{D}-1) s^2 - \tfrac{1}{4} \hat{D}\,  t u \big]
 - (-t)^{5/2}(2 s + 3 t) - (-u)^{5/2}(2 s + 3 u)\Big]  .\ \ \ \  \la{347}
\ee
Assuming that 2d momenta here are restrictions of  momenta of massless  3d particles, 
this  is indeed 
the same expression as  for the 1-loop  amplitude of  scattering 
of the   massless  3d modes  that   follows 
  directly from  the  action \rf{22}   expanded  near  the uncompactified  $\mathbb R^2$ membrane  vacuum (see \rf{3557}  below), modulo a factor of $(2 \pi R)^{-1}$ that comes from going from 3d to 2d.\foot{Since  $ T_2\sim$ (length)$^{-3}$ the amplitude in \eqref{347} has  dimension  (length)$^{-2}$ as appropriate for $d=2$, while the 3d amplitude in \rf{3557} has dimension (length)$^{-1}$  as appropriate for  $d=3$ (cf. \rf{a15},\rf{a7}), hence the need for an additional $R^{-1}$. This can also be recovered from the additional delta function in the S matrix, see \eqref{37} and \eqref{a15}. \label{f77}} 

In the  opposite  limit of $\rR \rightarrow 0$   the  form of the action  in \rf{33},\rf{36} 
implies that  only  the modes with $n=0$   should contribute, i.e. we should recover 
the scattering amplitude in the NG string  case. 
Taking this limit   directly in \rf{341}   with   $\Delta_s\to \rR^2 \Delta_s$ we find that the functions in \rf{342},\rf{343} 
have the following expansion
\begin{align}
F(1;\rR^2 \Delta_s) &= \frac{1}{ \rR^2 \Delta_s} + \frac{\pi^2}{3}+\OO(\rR)^2~, \quad \qquad 
F(0;\rR^2 \Delta_s) = -2 \log(2 \pi R \sqrt{\Delta_s}) + \OO(\rR)^2~, \\
F(-1;\rR^2 \Delta_s) &= \frac{1}{\pi^2}\zeta_R(3)+ \rR^2 \Delta_s \Big[2 \log(2 \pi \rR \sqrt{\Delta_s})-1\Big] + \OO(\rR)^4~.\la{3351}
\end{align}
Adding the   contributions of the three channels 
 we get for the   $\rR\to 0$   limit of the total 1-loop  amplitudes
  (assuming  $t=0$,  $u=-s$  and   fixing the effective string tension 
  $\hat T= 2 \pi \rR T_2$  in \rf{32})\foot{As $ \hat T\sim$ (length)$^{-2}$ 
 these amplitudes have   dimension  (length)$^{-2}$  as appropriate for $d=2$.
 }
\begin{equation}\begin{aligned}\la{350}\rR \to 0: \ \ \ 
% \lim_{\rR \rightarrow 0}
& \qquad   \hat{A}^{(1)}= {1\ov \hat T^2} \Big[
\frac{ \zeta_R(3)}{8\pi^3 \rR^2} s^2- \frac{\hat D -24 }{192\pi }  s^3 \Big]~,\\
 & 
%\lim_{\rR \rightarrow 0} 
 \hat{B}^{(1)} = \frac{i}{16 \,\hat T^2} s^3, \qquad \qquad \ \ 
 % \lim_{\rR \rightarrow 0}
  \hat{C}^{(1)} = {1\ov \hat T^2} \Big[ \frac{ \zeta_R(3)}{8\pi^3 \rR^2}  s^2 
  + \frac{\hat D -24 }{192\pi} s^3\Big]
   \, .\ \ \ \ \ \ \ 
\end{aligned}
\end{equation}
The  same  expression  is found by starting   with the first representation \rf{3355a}--\rf{3355c} 
 for the amplitudes in terms of the sums over $n$  where the  singular 
 $1\ov \rR^2$ terms
 appear  in  an equivalent form as 
 \be  \la{351}
  -\frac{ \zeta'_R(-2)}{2\pi  \rR^2}  s^2 \ , \qquad \qquad  \zeta_R'(-2) =- {1\ov 4 \pi^2} \zeta_R(3)\ . 
 \ee
 This    singular  contribution   cancels   once one  adds 
 the tadpole diagram  contribution in \rf{6643} (with $t=0$ and the  dependence on $\rR$ restored). 
 The remaining  finite parts of the amplitudes in \rf{350} are then  indeed the same as 
 in the NG   string case  in  \rf{235}.

\iffa The presence of the  extra 
 singular $1\ov \rR^2$   contributions in  $ \hat{A}^{(1)}$  and $ \hat{C}^{(1)}$
 appears to be  due to a non-commutativity of the  analytic continuation  using $\zeta_E$   in $F(-1; c)$ in \rf{338},\rf{bb7}  and  $c\to 0$ limit. 
 Note that  this singular terms 
have the same quartic in momenta 
  structure as the tadpole  contribution in \rf{6643}    and thus 
may be,  in principle, cancelled by a local counterterm.  
\fi

\subsection{Scattering amplitudes on uncompactified 
  membrane } \label{s3.4}

Let us   now consider the   S-matrix on  uncompactified   membrane
found by expanding the  Dirac  action \rf{21},\rf{22}  near   the plane  $\RR^{2}$ 
membrane vacuum. The scattering modes are then  massless 
3d fields representing  $\hat D = D-3$   transverse  membrane  coordinates.

The tree-level amplitudes are given  by the same expressions  \eqref{29}
as in the string case but
  with the 3d kinematic variables  subject  only  to the $s+t+u=0$  constraint.
 Including  the factor of the inverse membrane tension we thus have  (cf. footnote \ref{f77})
 \begin{equation}\la{2999}
 %v4-sign
d=3: \qquad \ \  \ A^{(0)} =- \frac{1}{2\, T_2}\,  tu~,\ \ \  \qquad B^{(0)} =- \frac{1}{2\, T_2}\,  su~,\ \ \  \qquad C^{(0)}=- \frac{1}{2\,T_2}\,  st~.
\end{equation}
The 1-loop 
amplitudes are  given   by  the same  integrals 
as in  \eqref{215}.
 In particular, the numerator $N_s$ in \rf{288},\eqref{2199} is the same
 as in the  string case, and so is the expression \eqref{221} with the  coefficients  as  in \eqref{223} and \eqref{2233} but  now with  $d=3$. 
 
 For $d=3$   the integral in \rf{221} is UV finite and we find that 
 \begin{align}
A^{(1)} &= \frac{1}{256\,T_2^2} \Big[ (-s)^{3/2} \Big((\tfrac{3}{32}  \hat{D}-1) s^2 - \tfrac{1}{4} \hat{D} t u \Big) +  (-t)^{3/2} (2 st + 3 t^2) +   (-u)^{3/2} (2 su + 3 u^2) \Big],\qquad\qquad\no   \\
B^{(1)} &= \frac{1}{256\,T_2^2} \Big[ (-t)^{3/2}\Big ( (\tfrac{3}{32} \hat{D}-1) t^2 - \tfrac{1}{4} \hat{D} su \Big) +   (-s)^{3/2} (2 ts + 3 s^2) +   (-u)^{3/2} (2 tu + 3 u^2) \Big] , \la{3557}\\
C^{(1)} &= \frac{1}{256\,T_2^2} \Big[ (-u)^{3/2}\Big( (\tfrac{3}{32} \hat{D}-1) u^2 - \tfrac{1}{4} \hat{D} st \Big) +   (-t)^{3/2} (2 ut + 3 t^2) +   (-s)^{3/2} (2 us + 3 s^2) \Big].\no 
\end{align}
As  $\sqrt{-s}=e^{i  {\pi\ov 2}} \sqrt{ |s|}  =i  \sqrt{ |s|} $  this 1-loop 
 4-point amplitude contains  an imaginary  part consistent with unitarity.\foot{Note that the amplitude is  non-zero  even for $\hat D=  D-3=0$  when there are no   transverse excitations  as  we formally assume that the propagator  is   always  normalized  to one   and  the interaction vertices are $D$-independent.}

While this  1-loop  S-matrix  is  finite, the non-renormalizability of the membrane action means that 
log UV divergences  may appear 
at higher loops. It would be very interesting  to  find a higher loop generalization of 
\rf{3557}  that  could be   the analog of  the pure-phase S-matrix \rf{245} in the NG string case
%\foot{We are grateful to S. Giombi for  this  suggestion.}
and, especially,     its   supermembrane \ci{Bergshoeff:1987cm} analog that may be better defined.

%%%%%%%%%%%%%%%%%%%%%%%%

\section{Integrable 
 \texorpdfstring{$T\bar{T}$}{TTbar} deformation of   infinite  tower of  2d fields }\la{s4}
%%%%%%%%%%%%%%%%%%%%%%%%%%%%%%

The integrability of the  NG  string  model 
 \eqref{22} has a natural explanation  in terms of a $T\bar{T}$ deformation of a free  2d scalar theory \cite{Cavaglia:2016oda,Bonelli:2018kik}. 
 While  the  effective 2d   model   obtained from compactified  membrane
  was  found  above  to be  non-integrable, 
  it is  possible   to  construct  an integrable   model   with the same free-field spectrum 
  by applying  a $T\bar{T}$ deformation  to the  massless plus massive tower of 2d fields  in \rf{33}. 
 In this case the  interaction Lagrangian will be different  from \rf{36}
 and will   not correspond to a compactification of  some  local Lorentz invariant theory in 3 dimensions.

\subsection{\texorpdfstring{$T\bar{T}$}{TTbar} deformation of free massless 2d  field}

The $T \bar{T}$ deformation can be applied to any 2d theory with an energy momentum tensor $T_{\a\b}$. When the seed theory is 
 integrable, its  $T\bar{T}$ deformation  will preserves integrability.
In general, the  ``$T\bar{T}$"  operator is defined as $\det T_{\mu\nu}$, i.e. 
\begin{equation}
O_{T \bar{T}} = \frac{1}{2} \epsilon^{\a\b} \epsilon^{\g\r} T_{\a \g} T_{\b \r} = \epsilon^{\g\r} T_{0 \g} T_{1 \r}\ , \la{41}
\end{equation}
with $\epsilon^{01}=-\epsilon^{10}=1$ the antisymmetric Levi-Civita symbol. 
 The corresponding  action  is ($\tn$ is a  deformation  parameter) 
\begin{equation} \label{42} 
%v4-sign
S =- \int d^2 \s\, \sqrt{-g}\ \mathcal L(\tn) \ , \qquad \qquad 
\partial_\tn \mathcal L = O_{T \bar{T}} \ , 
\end{equation}
where,  using that
%v4 sign 
  $T_{\a\b} = \frac{2}{\sqrt{-g}} \frac{\delta S}{\delta g^{\a\b}} =  g_{\a\b} \mathcal L - 2 \frac{\partial \mathcal L}{\partial g^{\a\b}}$,  we have 
\begin{align}
\quad O_{T \bar{T}} = -\mathcal L^2 + 2 \mathcal L g^{\a\b} \frac{\partial \mathcal L}{\partial g^{\a\b}} + 2 \epsilon^{\a\b} \epsilon^{\rho \sigma} \frac{\partial \mathcal L}{\partial g^{\a \rho}}\frac{\partial \mathcal L}{\partial g^{\b \sigma}}. \label{43}
\end{align}
Let us   start with the Lagrangian of $\hat{D}$  free massless  fields
\begin{equation}\la{44}
\mathcal L_0 = \frac{1}{2}  \partial^\alpha X^j \partial_\beta X^j, \qquad\qquad  j=1, \dots, \hat{D}.
\end{equation}
The  first-order  ($t\to 0$)  deformation of the Lagrangian will  then be  
\begin{equation}
\mathcal L_1 = \mathcal L_0 +  \tn \, O_{T \bar{T}} 
= \frac{1}{2} \partial_\alpha X^j \partial^\alpha X^j   -
\frac{1}{2} \tn\,  \Big(\ha \mathcal X^\a_\a \mathcal X^\b_\b 
% \frac{1}{2} \partial_\alpha X^j \partial^\alpha X^j \partial_\beta X^k \partial^\beta X^k 
 -\mathcal X_{\a \b} \mathcal X^{\a\b}\Big)\ , \la{45}
\end{equation}
where   we follow the notation in \rf{25}, i.e.  $\mathcal X_{\alpha \beta} \equiv  \partial_\alpha X^j \partial_\beta X^j$.
The interaction term  here  has the same form  as the first term in the expansion of the NG action  in \rf{24}. In $d=2$ there are only two independent   invariants $\JJ_1$ and $\JJ_2$   among    $\JJ_n\equiv \Tr[\mathcal X^n]$  (for instance, $\JJ_3 = - \frac{1}{2} \JJ_1^3 + \frac{3}{2} \JJ_1 \JJ_2, \ \  \JJ_4 = -\frac{1}{2} \JJ_1^4 + \JJ_1^2 \JJ_2 + \frac{1}{2} \JJ_2^2$). 
This allows  one to  obtain  the  closed form of the deformed  Lagrangian as 
$\mathcal L(\tn,\JJ_1,\JJ_2)$. Using that 
\begin{align}\la{46}
&\frac{\partial \mathcal L}{\partial g^{\alpha \beta}} = \frac{\partial \mathcal L}{\partial \JJ_1} \partial_\alpha X^j \partial_\beta X^j + \frac{\partial \mathcal L}{\partial \JJ_2} \partial_\alpha X^j \partial_\rho X^j \partial^\rho X^k \partial_\beta X^k , 
\\
&O_{T \bar{T}} = - \mathcal L^2 + 2 \mathcal L \Big( \frac{\partial \mathcal L}{\partial \JJ_1} \JJ_1 + 2 \frac{\partial \mathcal L}{\partial \JJ_2} \JJ_2 \Big)-2 (\JJ_1^2-\JJ_2) \Big[ \Big(\frac{\partial \mathcal L}{\partial \JJ_1}\Big)^2 + 2(\JJ_1^2 - \JJ_2)\Big(\frac{\partial \mathcal L}{\partial \JJ_2}\Big)^2+2 \JJ_1 \frac{\partial \mathcal L}{\partial \JJ_1} \frac{\partial \mathcal L}{\partial \JJ_2}     \Big], \no 
\end{align}
we find that  $\mathcal L(\tn,\JJ_1,\JJ_2)$   satisfying  the flow equation in \rf{42}  is given by 
\begin{equation}
\mathcal L(t,\JJ_1,\JJ_2) =  \frac{1}{  \tn } \Big[ \sqrt{1+   \tn\,  \JJ_1 + \frac{1}{2}  \tn^2\,  (\JJ_1^2-\JJ_2)} - 1 \Big]
% \frac{1}{2\bbeta\,  t} 
  =  \frac{1}{  \tn}  \Big[ \sqrt{-\text{det}(\eta_{\alpha \beta} + \tn\,  \partial_\alpha X^j \partial_\beta X^j  %\mathcal X_{\alpha \beta}
  )}-1\Big]~.\la{47}
\end{equation} 
It  is thus  equivalent to the NG Lagrangian in the static gauge (cf. \rf{21},\rf{23},\rf{22}). 

\subsection{\texorpdfstring{$T\bar{T}$}{TTbar} deformation of   a  tower of 
 massless   and   massive 2d  fields}
%%%%%%%%%%%%%%%%%%%%%%%%%%%%

Let us now start with $\hat{D}$ free scalars in 3 dimensions  and 
assume that  the third direction is  a circle of radius $\rR$. 
The equivalent 2d theory  is represented by  the  tower of  2d fields
which is the free part in the  compactified   membrane  action in  \rf{32}, i.e. 
\begin{equation}
\mathcal L_0 = \sum_{n=-\infty}^{\infty} \frac{1}{2}\Big( \partial_\alpha X_n^j \partial^\alpha X_{-n}^j + m_n^2 X_n^j X_{-n}^j \Big), \qquad \qquad m_n = \frac{n}{\rR}\ . \la{48}
\end{equation}
Computing  the  corresponding $T \bar{T}$  operator  in \rf{43} 
we get for  the  deformed  Lagrangian 
  \begin{align}
&\qquad \qquad \qquad \qquad  \mathcal L=   \mathcal L_0 + \tn\,  O_{T \bar{T}}   + \OO(\tn^2) \ , \la{4155} 
  \\
%O_{T \bar{T}} 
&
\mathcal L =\mathcal L_0   -\frac{1}{2} \tn  \sum^{\infty}_{n,q=-\infty} \Big(c_2 \partial_\alpha X^j_n \partial^\alpha X^j_{-n} \partial_\beta X^k_{q} \partial^\beta X^k_{-q} +c_3\partial_\alpha X^j_n \partial^\beta X^j_{-n} \partial_\beta X^k_{q} \partial^\alpha X^k_{-q}  \no \\
&\qquad  \qquad   + \tilde{c} \, m_n^2 m_q^2 X_n^j X_{-n}^j X_q^k X_{-q}^k \Big)+ \OO(\tn^2) \ , 
\qquad \qquad  c_2 =  \frac{1}{2}, \quad c_3 = -1,\quad  \tilde{c}= {1\ov 2} \ . 
\la{415} \end{align}
Here the $n=q=0$ term is the same  as in \rf{45}. 
Compared to the  quartic interaction term in the compactified membrane action 
in \rf{36}  here we get  only a subsector of $(\del X)^4$ terms with $n_1=-n_2$ and $n_3=-n_4$, no $X^2 (\del X)^2$  terms and the $X^4$ term with $n_1=-n_2$ and $n_3=-n_4$. %   with an opposite sign. % {\bf WHY ????}. 

Note that in contrast to \rf{36} the interaction term in the   2d  Lagrangian \rf{4155}   given  by the product of 
the  components of the two 2d  stress tensors 
cannot be obtained   from a local 3d action:  instead of a single $\delta_{n_1+n_2, -n_3-n_4}$    
it contains  
$\delta_{n_1, -n_2}  \delta_{n_3, -n_4}$ 
that  may only  originate 
  from a double  integral over the 3rd $S^1$  direction.

This difference is important for understanding of   why the membrane theory  interaction  term in \rf{36}  does not  correspond to  an integrable theory  while 
the  $T\bar T$  one in \rf{4155}  may.  Indeed,  the  structure of \rf{36} with only a  single $\delta_{n_1+n_2, -n_3-n_4}$   does not prohibit 
particle  transmutation processes   when, e.g., 2 particles with mass $m$  scatter into particle with mass $m$ and $m'$. Such processes are, however, excluded by 
the presence of  $\delta_{n_1, -n_2}  \delta_{n_3, -n_4}$ in \rf{415}.

Below we shall check that the theory \rf{415} is indeed integrable by computing the tree-level and 1-loop contributions to the corresponding scattering amplitude.

\subsubsection{Tree-level 4-point amplitude}

Let   us define  the  scattering amplitude of 4 particles  with indices $(i, j; k, l)$,  mode numbers or   masses  $(n_1,n_2; n_3, n_4)$  and momenta $(p_1,p_2; p_3,p_4)$  as 
(we take into account  the  specific structure of \rf{415})\foot{We  set $\rR=1$
  and  also   ignore the effective  coupling $\tn$ factor in \rf{415}.}
\begin{equation} \label{411}
\begin{aligned}
\mathcal M_{ij,kl} &= % \Big(
A\,  \delta_{ij} \delta_{kl} \, \delta_{n_1,-n_2} \delta_{n_3,-n_4} + B\,  \delta_{ik} \delta_{jl} \, \delta_{n_1,n_3} \delta_{n_2,n_4} + C\,  \delta_{il} \delta_{jk}\,  \delta_{n_1,n_4} \delta_{n_2,n_3} % \Big) \\
%&\qquad \times \delta^{(2)} (p_1+p_2-p_3-p_4)
~.
\end{aligned}
\end{equation}
We find  from \rf{415}  the following  tree-level amplitudes
%v4 signs 
\begin{equation} 
\begin{aligned}
A\zz[p_1,p_2,p_3,p_4] &=-\Big[ 2 c_2 (p_1 \cdot p_2) (p_3 \cdot p_4) + c_3 (p_1 \cdot p_4)(p_2 \cdot p_3) + c_3 (p_1 \cdot p_3)(p_2 \cdot p_4) + 2 \tilde{c} n_1^2 n_3^2 \Big]  ~,  \\
B\zz[p_1,p_2,p_3,p_4] &= -\Big[2 c_2 (p_1 \cdot p_3) (p_2 \cdot p_4) + c_3 (p_1 \cdot p_4)(p_2 \cdot p_3) + c_3 (p_1 \cdot p_2)(p_3 \cdot p_4) + 2 \tilde{c} n_1^2 n_2^2  \Big] ~, \la{412}\\
C\zz[p_1,p_2,p_3,p_4] &=-\Big[ 2 c_2 (p_1 \cdot p_4) (p_2 \cdot p_3) + c_3 (p_1 \cdot p_2)(p_3 \cdot p_4) + c_3 (p_1 \cdot p_3)(p_2 \cdot p_4) + 2 \tilde{c} n_1^2 n_2^2  \Big] ~.
\end{aligned}
\end{equation}
Here  $B\zz[p_1,p_2,p_3,p_4]$ and $C\zz[p_1,p_2,p_3,p_4]$ can 
be of course obtained from $A\zz[p_1,p_2,p_3,p_4]$  by $(p_2 \leftrightarrow -p_3, n_2 \leftrightarrow -n_3)$ and $(p_2 \leftrightarrow -p_4, n_2 \leftrightarrow -n_4)$ respectively. 
In the massless scattering case of $n_1=n_2=n_3=n_4=0$ 
 one  recovers the expressions in \eqref{210}.
 
Expressing \rf{412} in terms of  the Mandelstam variables in \rf{a11}  we get 
\begin{equation}
\begin{aligned}\la{413}
%v4 signs
A\zz &= - \frac{1}{4} (2 c_2 + c_3) \big[s^2-2 s (n_1^2+n_3^2)\big] - 2 (c_2+c_3 + \tilde{c}) n_1^2 n_3^2+ \frac{1}{2} c_3 \big[t u - (n_1^2-n_3^2)^2  \big]~, \\
B\zz &=  -\frac{1}{4} (2 c_2 + c_3) \big[t^2-2 t (n_1^2+n_2^2)\big] - 2 (c_2+c_3 + \tilde{c}) n_1^2 n_2^2+ \frac{1}{2} c_3 \big[s u - (n_1^2-n_2^2)^2  \big]~, \\
C\zz &= - \frac{1}{4} (2 c_2 + c_3) \big[u^2-2 u (n_1^2+n_2^2)\big] - 2 (c_2+c_3 + \tilde{c}) n_1^2 n_2^2+ \frac{1}{2} c_3 \big[s t - (n_1^2-n_2^2)^2  \big]~.
\end{aligned}
\end{equation}
Here we took into account the delta-symbol constraints on the mode numbers or masses
 in  \eqref{411}.
 % i.e. $A[p_1,p_2,p_2,p_4] \delta_{n_1,-n_2} \delta_{n_3,-n_4} = A \delta_{n_1,-n_2} \delta_{n_3,-n_4}$, etc. 
For the specific coefficients in \eqref{415} the first two terms in each expression in \rf{413} vanish, i.e. 
\begin{equation}\la{414}
%v4 signs
A\zz =- \frac{1}{2} \big[t u - (n_1^2-n_3^2)^2\big]
~, \qquad B\zz =- \frac{1}{2} \big[s u - (n_1^2-n_2^2)^2\big]~, \qquad C\zz = -\frac{1}{2} \big[ s t - (n_1^2-n_2^2)^2\big]~.
\end{equation}
One can check using \rf{a11}--\rf{a6} % the expressions in appendix \ref{a2} 
that in the 2d case 
\begin{align}
 \big[t u -(n_1^2-n_3^2)^2\big] \delta_{n_1,-n_2} \delta_{n_3,-n_4}  = \big[s t -(n_1^2-n_2^2)^2\big] \delta_{n_1,n_4} \delta_{n_2,n_3} =0 \ \ 
 \ %\qquad
 \rightarrow \ \
 A\zz=C\zz=0,\no \\
\Big(s u - 4 n_1^2 n_2^2 + \big[s-(n_1^2+n_2^2)\big]^2\Big) \delta_{n_1,n_3} \delta_{n_2,n_4} = 0 \ \rightarrow \ \
%v4 sign
B\zz = -2 n_1^2 n_2^2 + \frac{1}{2} \big[s-(n_1^2+n_2^2) \big]^2.\la{416}
\end{align}
Thus the  only  non-vanishing amplitude is the transmission $B$  one, consistent with the  integrability. 

For the scattering  of massive particles  we may  use the notation in terms of the 
rapidities, $\omega_r = \sqrt{\vec p_r{}^2 + n_r^2} = n_r \cosh \theta_r$ and $\vec p_r= n_r \sinh \theta_r$, so that the  
amplitude
$B\zz$  in \rf{416} 
 may be written as 
\begin{equation}
%v4 sign
B\zz = 2 n_1^2 n_2^2 \sinh^2(\theta_1-\theta_2)~.  \la{4164}
\end{equation}
We conclude that  the  tree-level 4-point 
S-matrix is proportional to the identity and the Yang-Baxter equation is trivially satisfied, 
in agreement   with the  expected integrability  of the $T\bar{T}$ deformation.

\subsubsection{1-loop scattering amplitude for massless   fields} 

To compare to the results in section 3 for the 1-loop massless  scattering  amplitude  in the compactified membrane theory   let us consider the same problem in the case of the $T\bar T$ deformed   theory \rf{4155}. We shall thus assume that 
for  the external particles $n_1=n_2=n_3=n_4=0$. The mode number in the loop 
will be denoted as $n$. 

Let us   start with the contribution of the  bubble diagrams  in Figure 1. 
The resulting expression for $A^{(1)}_s$  amplitude  is given by 
the same  expressions as in \rf{319},\rf{317},\rf{288}   where   instead of \rf{320}  we  get (in the $s$-channel) 
\begin{align}
& N_0 = 0~, \qquad N_2 = -s^3 x(1-x) \delta_{n,0}~, \qquad N_4 = \frac{1}{4}(\hat{D}-4 \delta_{n,0}) s^2~,\no  \\ \quad 
& M_{\mu \nu} = (\hat{D}-2) s \big(p_{1,\mu} p_{2,\nu}  + p_{3,\mu} p_{4,\nu}    \big) ~, \qquad N_{\mu \nu \rho \sigma} = 4 \hat{D} p_{1,\mu} p_{2,\nu} p_{3,\rho} p_{4,\sigma} ~, \no \\
& \begin{aligned} N_{\mu \nu} &=  \big[-2 s^2 x(1-x) + s (\hat{D}-2) n^2\big]
\big(p_{1,\mu} p_{2,\nu}  + p_{3,\mu} p_{4,\nu}    \big)~.\la{418}
\end{aligned}
\end{align}
This  leads to \rf{224}  where now  (cf. \rf{3223}) 
\begin{equation}
\begin{aligned}
\GG_{0,s} &= \frac{1}{4} s^2 \big[
n^2- x(1-x) s\big]  \Big[\hat{D}\big[n^2-x(1-x) s\big] + 8 x(1-x) s \delta_{n,0}\Big]~, \\
\GG_{2,s} &= -s^3 x(1-x) \delta_{n,0}~, \qquad \GG_{4,s} = -\frac{1}{2} \hat{D} tu + s^2 \delta_{n,0}~.\la{417}
\end{aligned}
\end{equation}
For  the massless loop  mode 
 $n=0$ this reduced to the coefficients found in  \eqref{2255}. 
 
 Similarly, in the  $t$-channel   we get  (the $u$-channel expressions are found  by   $t \leftrightarrow u$ and $p_3 \leftrightarrow p_4$)
\begin{align}
& N_0 = \frac{1}{2} t^4 x^2 (1-x)^2 \delta_{n,0}~, \quad N_2 = t^3 x (1-x) \delta_{n,0}~, \quad N_4 = \frac{1}{2} t^2 \delta_{n,0}~, \quad M_{\mu \nu} = 0 ~, \quad N_{\mu \nu \rho \sigma} = 0\qquad  \no \\
&N_{\mu \nu} =  -2 t^2 x (1-x) \big(
p_{1,\mu} p_{4,\nu} - p_{1,\mu} p_{2,\nu} - p_{3,\mu} p_{4,\nu}  + p_{3,\mu} p_{2,\nu}
\big)  - t^2 \big(p_{1,\mu} p_{2,\nu}  + p_{3,\mu} p_{4,\nu}    \big). \la{419}
\end{align}
This  gives
\begin{equation} 
\begin{aligned}\la{420}
\GG_{0,t} &= 2 t^4 x^2(1-x)^2 \delta_{n,0}, &\qquad \GG_{2,t}&=\big[\frac{1}{2} t^2 s + 3 t^3 x(1-x) \big] \delta_{n,0}, &\qquad \GG_{4,t} &=0, \\
\GG_{0,u} &= 2 u^4 x^2(1-x)^2 \delta_{n,0}, &\qquad \GG_{2,u}&=\big[\frac{1}{2} u^2 s + 3 u^3 x(1-x) \big] \delta_{n,0}, &\qquad \GG_{4,u}&=0.
\end{aligned}
\end{equation}
As a result,   the  pole part of the total $A^{(1)}$   amplitude reads
\begin{equation}
A^{(1)}_{n,\epsilon} = \frac{1}{96 \pi} \Big( \frac{1}{\epsilon} - \gamma + \ln 4 \pi \Big) \Big[ - (\hat{D}-6 \delta_{n,0}) s t u + 6 \hat{D} n^2 t u \Big].\la{421}
\end{equation}
Compared to the compactified membrane case  in \rf{325} here we get  no $s^2$ term   so this 
expression  vanishes  for $t=0$  choice of the 2d   kinematics  even   before summation over $n$, i.e.  the amplitude is UV finite  for fixed $n$. 

 For the remaining finite part of the  fixed-$n$ amplitudes 
  we find (setting $t=0,\,  u=-s$)\foot{To  find the expression for  $B^{(1)}_{n,f}$ we used the small $n$ expansion of the function $Q_n(s)$ in \rf{3322},\rf{231} implying that 
$ \big[Q_n(s)+Q_n(-s)\big]\delta_{n,0} = {2 i \pi}{s}^{-1} \delta_{n,0}.$}
\begin{align}
A^{(1)}_{n} &= -\frac{1}{192 \pi} \Big[ (\hat{D}-24 \delta_{n,0}) s^3  - 6\hat{D} n^2 s^2  \Big],\la{422} \\
B^{(1)}_{n} &= \frac{1}{16} \Big[ i   s^3  \delta_{n,0} -  \frac{1}{\pi} \hat{D} n^2( 1 - \ln n^2)\, s^2 \Big] ,\la{423} \\
C^{(1)}_{n} &= \frac{1}{192 \pi} \Big[ (\hat{D}-24 \delta_{n,0}) s^3  + 6\hat{D}  n^2s^2   \Big].\la{424}
\end{align}
For $n=0$ this of course reduces to the string scattering amplitudes in \rf{235}.

The total amplitude is given   by the sum over $n$    which we may define again using the Riemann 
$\zeta$-function  prescription as in \rf{326}.  Then the  $\hat D s^3$   and $\hat D s^2 n^2$ terms in
$A^{(1)}$  and $C^{(1)}$ disappear  and the remaining real $s^3$ terms  
 can be  cancelled   by  a  local counterterm as in \rf{239}. 
 The real  contribution to $B^{(1)}_{n} $   proportional  to $\zeta_R'(-2) \hat D  s^2$  (cf. \rf{335}) 
 should   cancel against the contribution of the tadpole diagrams  as in \rf{6643}.\foot{The direct computation of the tadpole diagram   requires  
 fixing the 6-point vertex in the deformed Lagrangian in  \rf{4155}.}
 
 Then the final expression for  the 1-loop massless amplitude is the same   as in the purely massless (string) 
 theory. This is  consistent with the origin of \rf{415}  as 
 a $T\bar{T}$ deformation of  the free (integrable)   model:
   the S-matrix should  be given by that of the undeformed theory, dressed by a CDD factor. 
   The CDD factor is only sensitive to the quantum numbers  of the external particles, not those  of the virtual 
   particles in the loop. 
  Therefore, we  should  indeed end up with  the   same amplitude  as in the purely massless case. 

Let us note that in the  case of   1-loop  correction to 
scattering of massive fields in the  deformed theory \rf{4155} 
one will   find UV  divergent  terms that need to be cancelled  by 
 appropriate counterterms  \cite{Rosenhaus:2019utc}.
The   ambiguity in the structure of  finite  counterterms   may be  fixed by 
requiring  that  the theory should be 
 integrable at the   quantum level, i.e. its  S-matrix  should 
 satisfy the YB  equation.\foot{This 
was demonstrated on an example of a $T\bar T$ deformation  of a
 single massive field 
 in \cite{Rosenhaus:2019utc}  using  a momentum  cutoff regularization. We have checked that 
 similar result  is found  using dimensional regularisation.}

%%%%%%%%%%%%%%%%%%%%%%%%%%
\section{Concluding remarks}\la{s5}
%%%%%%%%%%%%%%%%%%%%%%%%

We have   found  that the  S-matrix  of the  effective 2d model    corresponding to the 
 bosonic membrane  action  expanded near  the   cylindrical  vacuum 
 is not integrable,  so the question of  possible hidden symmetries  in membrane theory 
 remains open. 
%The non-integrability   should apply also to the   supermembrane case but   it is likely  that the corresponding scattering amplitudes should have much simpler   structure.  

There are  several possible extensions. 
An  interesting  problem is to extend  the  computation of the    S-matrix 
in the  bosonic   string and membrane theories
 to the Green-Schwarz   superstring\foot{In the integrable 
 superstring case  tree-level   S-matrix    was already discussed  in \ci{Cooper:2014noa}.}
  and the supermembrane  theories.
 Starting with the GS      string in the static gauge one finds 
\ci{Roiban:2007jf}  that its  partition function is UV finite   and  trivial   at 1-loop   and also  2-loop   orders 
(the  2-loop  log divergences  cancel also in \adss  case  \ci{Roiban:2007jf,Giombi:2009gd}). 
It would   be  important to  show  that the 
 corresponding 2d   scattering amplitudes are  also 1-loop and 2-loop  finite, despite formal non-renormalizability of the 
 GS theory. 
 
  Similarly, in the  case of the   supermembrane in flat target space 
     it would  be interesting   to check  first   that 
  the 2-loop   partition function of  the cylindrical membrane 
    is UV   finite (the 1-loop  one is  always finite in $d=3$). 
One may then   compute the  corresponding 
S-matrix  to 2-loop order to see if it is also   well defined   and have a simple structure. 

One may  also  investigate the  $T\bar T$  deformation of the free 
2d    theory  obtained   from compactified   supermembrane 
 (i.e.   containing  free  superstring  modes  plus a  tower of massive   2d fields).  
 This should produce an integrable model  with   an infinite set of 2d  bosons and fermions, 
 generalizing the static-gauge GS action.

%extensions -- add fermions in 3d  in flat space.

% consider M2 in ads4 x s7 and consider  S-matrix
 % in BMN limit -- generalization of BMN  S-matrix in 10d ? 
  %  connection to ABJM -- generalization of   S-matrix ? 

\section*{Acknowledgements}
We  are grateful to S. Giombi  for the   initial suggestion to study  
 S-matrix in membrane theory  and  useful  discussions. 
 We also thank 
  D. Polvara, R. Roiban  and A. Sfondrini  for  helpful comments  and R. Metsaev  for important remarks   on the draft. 
FS  is supported  by  the European Union  Horizon 2020 research and innovation programme under the Marie Sklodowska-Curie grant agreement number 101027251, and would like to thank the participants of the Filicudi workshop on Integrability in lower-supersymmetry systems for stimulating discussions.
AAT is supported by the STFC grant ST/T000791/1. 
He also acknowledges
the hospitality of  Nordita program ``New perspectives on quantum field theory with boundaries, impurities and defects" 
at the  final stage of this work.

\

 %\newpage

%%%%%%%%%%%%%%%%%%%%%%%%%%%%%

%\newpage
%%%%%%%%%%%%%%%%%%%%%%%%%%%%%
\appendix
\section{Notation and  basic relations}\la{a1}

We use the Minkowski signature $(-+...+)$.  
$D$    denotes  the target space dimension of a string or membrane theory, 
$d$ is the world-volume 
 dimension. 
 We  also   use the notation $\hat{D}=D-d$  for 
  the number of  the physical ``transverse'' fields remaining after fixing a  static gauge.

%\subsection*{Mandelstam variables}

For the four-point scattering 
 involving particles with momenta $p_r$  and masses $m_r$ ($r=1,2,3,4$)
we  define  the Mandelstam variables as ($p_1,p_2$ are incoming and $p_3,p_4$ 
are outgoing momenta, $p_1+p_2=p_3+p_4$,\  $p_r^2 = - m^2_r$)
\begin{align}  %\label{a11}\begin{aligned}
s &= -(p_1 +p_2)^2 = -(p_3+p_4)^2 = m_1^2 + m_2^2 -2 p_1 \cdot p_2 = m_3^2 + m_4^2 - 2 p_3 \cdot p_4~, \no \\
t &= -(p_1 - p_3)^2=-(p_2-p_4)^2 = m_1^2 + m_3^2 + 2 p_1 \cdot p_3 = m_2^2 + m_4^2 + 2 p_2 \cdot p_4~,\no  \\
u &= -(p_1-p_4)^2 = -(p_2-p_3)^2 = m_1^2 + m_4^2 + 2 p_1 \cdot p_4 = m_2^2 + m_3^2 + 2 p_2 \cdot p_3~,\no \\
& s + t + u = - p_1^2 - p_2^2 - p_3^2 - p_4^2=m_1^2+m_2^2+m_3^2+m_4^2~ \ .\la{a11}
\end{align}
In two dimensions  the Mandelstam variables satisfy an additional constraint: 
we have $4 \times 2$ components  of momenta   subject to 2 energy-momentum conservation  and 4 mass shell constraints plus  there is 1  parameter of $SO(1,1)$ Lorentz transformation leaving only $8-2-4-1=1$ 
independent kinematic variable.  This constraint can be expressed, e.g.,  as follows: 
\begin{align}
0 &= -4 \left| 
\begin{matrix}
p_1 \cdot p_1 & p_1 \cdot p_2 & p_1 \cdot p_3 \\
p_2 \cdot p_1 & p_2 \cdot p_2 & p_2 \cdot p_3 \\
p_3 \cdot p_1 & p_3 \cdot p_2 & p_3 \cdot p_3
\end{matrix}
\right| = -4 \left| 
\begin{matrix}
-m_1^2 & -\frac{1}{2} (s-m_1^2-m_2^2) & \frac{1}{2} (t-m_1^2-m_3^2) \\
-\frac{1}{2} (s-m_1^2-m_2^2) & -m_2^2 & \frac{1}{2} (u-m_2^2-m_3^2) \\
\frac{1}{2} (t-m_1^2-m_3^2) & \frac{1}{2} (u-m_2^2-m_3^2) & -m_3^2
\end{matrix}
\right| 
\no \\
&  =  s t u + s (m_1^2+m_2^2)(m_3^2+m_4^2) + t (m_1^2+m_3^2)(m_2^2+m_4^2)+u (m_1^2 + m_4^2)(m_2^2+m_3^2)\no \\ %+C \ , \\
&\ \ \   -\frac{1}{6} \Big(\sum_{j=1}^4 m_j^2\Big)^3 - \frac{1}{2} \Big(\sum_{j=1}^4 m_j^2 \Big) \Big(\sum_{j=1}^4 m_j^4 \Big) + \frac{2}{3} \Big(\sum_{j=1}^4 m_j^6 \Big).\la{a12}
\end{align}
\iffa with constant
\begin{equation}
C = -\frac{1}{6} \Big(\sum_j m_j^2\Big)^3 - \frac{1}{2} \Big(\sum_j m_j^2 \Big) \Big(\sum_j m_j^4 \Big) + \frac{2}{3} \Big(\sum_j m_j^6 \Big).
\end{equation}\fi
Using     \eqref{a11}  allows to express, e.g., $t$ and $u$  in terms of $s$. 
 We choose the solution of the resulting quadratic equations  that
  becomes  $t=0$ and $u=-s$  in  massless  case\foot{The other solution corresponds to $u=0$ and $t=-s$  and is simply  obtained  by $t \leftrightarrow u$.}
\begin{align}\la{a6}
t &=  \frac{m_1^2+m_2^2+m_3^2+m_4^2 -s}{2} - \frac{(m_1^2-m_2^2)(m_3^2-m_4^2)-\Sigma(s)}{2 s} , \no \\
u &= \frac{m_1^2+m_2^2+m_3^2+m_4^2 -s}{2} + \frac{(m_1^2-m_2^2)(m_3^2-m_4^2)-\Sigma(s)}{2 s},\\
& \Sigma(s)  \equiv  \sqrt{(s-v_1)(s-v_2)(s-v_3)(s-v_4)},\no \\
& v_1 = (m_1-m_2)^2, \qquad v_2=(m_1+m_2)^2, \qquad v_3=(m_3-m_4)^2, \qquad v_4=(m_3+m_4)^2.\no 
\end{align}

In general, the  S-matrix can be represented 
 as  
 \be \S=1+i  \rT \ , \la{a44}\ee
 where $1$ denotes the identity (corresponding to the free theory) and $\rT$  encodes the contribution of  interactions.
  The operators  $1$ and $\rT$ act on the asymptotic states for the 4-point scattering as
\begin{align} \label{a14}
\big< X^k(\vec{p}_3) X^l(\vec{p}_4) | 1 | X^i(\vec{p}_1) X^j(\vec{p}_2) \big> &= \delta^{(d-1)}(\vec{p}_1 - \vec{p}_3) \delta^{(d-1)}(\vec{p}_2-\vec{p}_4)\, \delta^{ik}\delta^{jl} , \\ \label{a15}
\big< X^k(\vec{p}_3) X^l(\vec{p}_4) | \TTT | X^i(\vec{p}_1) X^j(\vec{p}_2) \big> &=\mathcal  M^{ij,kl}[p_1,p_2,p_3,p_4]
\,  \delta^{(d)}(p_1+p_2-p_3-p_4) \,  \prod_{r=1}^4 \frac{1}{\sqrt{2\omega_r}}~.
\end{align}
Here $p=(\omega,\vec{p})$ with $\omega$ being  the energy and $\vec{p}$ denoting the spatial 
components of the momentum (we shall use this notation also  in $d=2$). 
The scattering amplitude  may be written as 
\begin{equation}\la{a7}
\mathcal M_{ij,kl}[p_1,p_2,p_3,p_4] = A[p_1,p_2,p_3,p_4]\,  \delta_{ij} \delta_{kl} + B[p_1,p_2,p_3,p_4]\,  \delta_{ik} \delta_{jl} + C[p_1,p_2,p_3,p_4]\,  \delta_{il} \delta_{jk}~,
\end{equation}
where it  is assumed that 
all  particles are on shell, i.e.  one has 
$\omega_r = \sqrt{\vec{p}_r{}^2 + m_r^2}$.

 Note that $\mathcal M_{ij,kl}$ has mass dimension $4-d$, so that 
 the right-hand sides of both \eqref{a14} and \eqref{a15} have the same 
 mass dimension $2-2d$. 
 This is consistent with the  expressions for the amplitudes  in the main text. 
 Restoring the tension factors   both tree-level and one-loop amplitudes 
  $\mathcal M_{ij,kl}$ have  mass dimension 2 in $d=2$  and 1 in  $d=3$.

Specialising to $d=2$, 
if  the theory is  integrable, then there is no particle transmutation, i.e. 
 only the processes with $(\vec{p}_1,m_1)=(\vec{p}_3,m_3)$ and $(\vec{p}_2, m_2)=(\vec{p}_4,m_4)$ are allowed.\footnote{Here we  are assuming
 that  $\vec{p}_1 > \vec{p}_2$ and $\vec{p_3} > \vec{p}_4$  ($\vec{p}$ is simply a scalar in 2d). 
  For the  opposite ordering of the outgoing momenta, only the processes with $(\vec{p}_1,m_1)=(\vec{p}_4,m_4)$ and $(\vec{p}_2, m_2)=(\vec{p}_3,m_3)$ are allowed.} It is then customary to pull out an overall $\delta(\vec{p}_1-\vec{p_3}) \delta(\vec{p_2}-\vec{p}_4)$ factor in \rf{a14},\rf{a15} and define the S-matrix element as
 \begin{align}\la{a8}
\S_{ij,kl}(\vec{p_1},\vec{p}_2) &= \delta_{ik} \delta_{jl} + i  \frac{J(\omega_1,\omega_2)}{4 \omega_1 \omega_2} \mathcal M_{ij,kl}[\vec{p}_1,\vec{p}_2,\vec{p}_1,\vec{p}_2]~, \qquad  J(\omega_1,\omega_2)= \frac{\omega_1 \omega_2}{|\vec{p}_1 \omega_2 - \vec{p}_2 \omega_1|}~.
\end{align}
%%%%%%
\iffa 
%   \subsection*{Scattering matrix and amplitudes}\la{a2}
The 4-point   
S-matrix  in $d=2$  is related to the scattering amplitude  as %through the formula (here for $d=2$)
\begin{align}
& {\rm S} = 1 + i \TTT, \ \ \ \ \ \ \ 
 i \TTT_{ij,kl}[p_1,p_2,p_3,p_4] = {\cal M}_{ij,kl}[p_1,p_2,p_3,p_4] \, 
\prod_{j=1}^4  \frac{1}{\sqrt{2\omega_j }} \ ,  \la{a77}   \\
& {\cal M}_{ij,kl} = M_{ij,kl}[p_1,p_2,p_3,p_4] \ \delta^{(d)} (p_1 + p_2 -p_3 - p_4) \ , \la{a7}
\end{align}
where  the relativistic dispersion relation is 
$\omega_j = \sqrt{\vec{p}_j^2 + m_j^2}$ (we use $\vec{p}_j$   to denote spatial $(d-1)$-component momenta). 
Note that as  $1$   in S   contains  $\delta$-function,  $\TTT$  has mass dimension $d$   and thus $M_{ij,kl}$ then has  mass dimension 
$2$. 
Specialising to $d=2$, 
if  the theory is  integrable, then there is no particle transmutation, i.e. 
 only the processes with $m_1=m_3$ and $m_2=m_4$, or $m_1=m_4$ and $m_2=m_3$ are allowed.
The  energy-momentum conservation $\delta$-function then reads 
 \begin{align}\la{a8}
\delta^{(2)}(p_1 &+ p_2 - p_3 - p_4) = \delta(\vec{p}_1 + \vec{p}_2 -\vec{p}_3 - \vec{p}_4)\   \delta(\omega_1+\omega_2-\omega_3-\omega_4)\no  \\
&=\frac{\omega_1 \omega_2}{|\vec{p}_1 \omega_2 - \vec{p}_2 \omega_1|} \ \Big[\delta(\vec{p}_1 - \vec{p}_3) \delta(\vec{p}_2 - \vec{p}_4) + \delta(\vec{p}_1 - \vec{p}_4) \delta(\vec{p}_2 - \vec{p}_3) \Big]\ . \qquad 
\end{align}
%\end{equation}
\fi
%%%%
%%%%%%%%%%%%%%%%%%%%%%%%%%%%%%%%
Let  us also recall  that  a necessary condition  for  integrability of a 2d theory 
 is that the S-matrix satisfies the quantum Yang-Baxter equation that has the following operator  form 
\begin{equation}\la{a9}
\S_{12}\, \S_{13}\, \S_{23} = \S_{23}\, \S_{13}\, \S_{12}  \ . 
\end{equation}
This relation  is automatically satisfied  if  $\S$ is proportional to the identity.  
Using \rf{a44}  we then get to  the leading interaction order 
\begin{equation}\la{a10}
\com{\rT_{12}}{\rT_{13}} +\com{\rT_{13}}{\rT_{23}} +\com{\rT_{12}}{\rT_{23}} =0\ . 
\end{equation}

\section{Some useful  integrals}
\label{a2}

To compute 1-loop momentum   integrals one may use, e.g.,    Feynman or Schwinger  parametrization 
\begin{align}
 \frac{1}{A_1 \dots A_n} & =  \int_0^1 d x_1 \dots d x_n\,  \frac{(n-1)!}{(x_1 A_1 + \dots + x_n A_n )^n} \, \delta(x_1 + \dots +x_n-1)\ ,\no  \\
&= \int_0^\infty d\ss_1 \dots \int_0^\infty  d\ss_n\  e^{-(\ss_1 A_1 + \dots +\ss_n A_n)}\ , \\
&  
x_i = \frac{\ss_i}{\ss_1 + \dots +\ss_n}, \qquad \qquad i=1,\dots,n\no \ . 
\end{align}
The standard momentum integrals in $d$ dimensions are 
\begin{align}
I_0 = \int \frac{d^dp}{(2 \pi)^d} \frac{1}{(p^2 + \Delta)^n} &= \frac{i\Gamma(n-\frac{d}{2})}{(4 \pi)^{d/2} \Gamma(n)}  \Big(\frac{1}{\Delta} \Big)^{n-\frac{d}{2}}~,\no \\
I_2=\int \frac{d^dp}{(2 \pi)^d} \frac{p^2}{(p^2 + \Delta)^n} &= \frac{i\Gamma(n-1-\frac{d}{2})}{(4 \pi)^{d/2} \Gamma(n)} \Big(\frac{1}{\Delta} \Big)^{n-1-\frac{d}{2}} \frac{d}{2}~,\no  \\
I_2^{\mu \nu}=\int \frac{d^dp}{(2 \pi)^d} \frac{p^\mu p^\nu}{(p^2 + \Delta)^n} &=  \frac{i\Gamma(n-1-\frac{d}{2})}{(4 \pi)^{d/2} \Gamma(n)}  \Big(\frac{1}{\Delta} \Big)^{n-1-\frac{d}{2}} \frac{1}{2} \eta^{\mu \nu}~, \la{b1}\\
I_4=\int \frac{d^dp}{(2 \pi)^d} \frac{(p^2)^2}{(p^2 + \Delta)^n} &=  \frac{i\Gamma(n-2-\frac{d}{2}) }{(4 \pi)^{d/2} \Gamma(n)} \Big(\frac{1}{\Delta} \Big)^{n-2-\frac{d}{2}}\frac{d(d+2)}{4} ~, \no \\
I_4^{\mu \nu \rho \sigma}=\int \frac{d^dp}{(2 \pi)^d} \frac{p^\mu p^\nu p^\rho p^\sigma}{(p^2 + \Delta)^n} &=  \frac{i\Gamma(n-2-\frac{d}{2})}{(4 \pi)^{d/2} \Gamma(n)}  \Big(\frac{1}{\Delta} \Big)^{n-2-\frac{d}{2}} \frac{1}{4} \Big(\eta^{\mu \nu} \eta^{\rho \sigma} + \eta^{\mu \rho} \eta^{\nu \sigma} + \eta^{\mu \sigma} \eta^{\nu \rho} \Big)~, \no 
\end{align}
where  in general $\int d^d p \, p^\mu p^\nu f(p^2) = \frac{1}{d} \eta^{\mu \nu} \int d^d p\,  p^2 f(p^2)$, etc.

The 1-loop  integrals we use are \rf{b1}  with  $n=2$  and
to define them  we   apply   dimensional regularization with 
$d=2-2 \epsilon$.  Taking into account that 
$\Gamma(-k +\epsilon) = \frac{(-1)^k}{k!} \big(\frac{1}{\epsilon} - \gamma+ \sum_{r=1}^k \frac{1}{r} \big) +O(\epsilon)~,
$
with $\gamma$ the Euler-Mascheroni constant, 
we get  for the above integrals 
\begin{align}
I_0 &= \frac{i}{4 \pi} \frac{1}{\Delta}~, \qquad \qquad 
I_2 = \frac{i}{4 \pi} \Big( \frac{1}{\epsilon} - \gamma + \ln 4 \pi - \ln \Delta \Big) ~, \no\\
I_2^{\mu \nu} &= \frac{i}{4 \pi} \Big( \frac{1}{\epsilon} - \gamma + \ln 4 \pi  - \ln \Delta \Big) \frac{1}{2} \eta^{\mu \nu} ~, \quad \ \ 
I_4 = \frac{i}{4 \pi} \Big( \frac{1}{\epsilon} - \gamma + \ln 4 \pi  -\ln \Delta +1\Big)(-2 \Delta) ~,\la{b2} \\
I_4^{\mu \nu \rho \sigma} &=  \frac{i}{4 \pi} \Big( \frac{1}{\epsilon} - \gamma + \ln 4 \pi -\ln \Delta +1\Big) (-\Delta) \frac{1}{4} \Big(\eta^{\mu \nu} \eta^{\rho \sigma} + \eta^{\mu \rho} \eta^{\nu \sigma} + \eta^{\mu \sigma} \eta^{\nu \rho} \Big)~.\no
\end{align}

In section 3  we  use 
 the Epstein $\zeta$-function  defined by   (assuming  $c>0$):
\begin{equation} \begin{aligned}
\zeta_E(w;c) &= \sum_{n=-\infty}^{\infty} \frac{1}{(n^2+c)^w} = \frac{1}{\Gamma(w)} \sum_{n=-\infty}^{\infty} \int_0^\infty dy \, y^{w-1} e^{- (n^2+c)y} \ . \la{b4}
\end{aligned}
\end{equation}
Applying the Poisson resummation $\sum_{n=-\infty}^{\infty} e^{-y n^2} =( \frac{\pi}{y})^{1/2} \sum_{n=-\infty}^{\infty} e^{-\frac{\pi^2 n^2}{y}}$
\iffa 
At this point we use the fact that 
\begin{equation}
\sum_{n=-\infty}^{\infty} e^{-y n^2} = \Big( \frac{\pi}{y} \Big)^{1/2} \sum_{n=-\infty}^{\infty} e^{-\frac{\pi^2 n^2}{y}}
\end{equation}\fi 
and assuming the  integral and the sum commute  we get 
\begin{equation} \begin{aligned}
\zeta_E(w;c) &= \frac{\sqrt{\pi}}{\Gamma(w)} \int_0^\infty dy \, y^{w-\frac{3}{2}} e^{- c\,y} \sum_{n=-\infty}^{\infty} e^{-\frac{\pi^2 n^2}{y}} \\
&= \frac{\sqrt{\pi}}{\Gamma(w)} \int_0^\infty dy \, y^{w-\frac{3}{2}} e^{- c\,y} +  \frac{2\sqrt{\pi}}{\Gamma(w)} \int_0^\infty dy \, y^{w-\frac{3}{2}} e^{- c\,y}  \sum_{n=1}^{\infty} e^{-\frac{\pi^2 n^2}{y}} \ , \la{b6}
\end{aligned}
\end{equation}
where in the last equality we separated  the  $n=0$  contribution.  We thus find that 
\begin{equation}\la{b7}
F(w;c)\equiv 
\Gamma(w)\, \zeta_E(w;c)
 ={ \sqrt{\pi} \ov c^{w-\ha} } \Gamma(w-\tfrac{1}{2} )+ {4 \pi^w\ov  (\sqrt{c})^{w- \ha}} 
 \sum_{n=1}^\infty n^{w-\ha } K_{w- \ha }(2 \pi n \sqrt{c})\ , 
\end{equation}
where $K_\nu=K_{-\nu}$ is a  Bessel function, see, e.g.,~\cite{Elizalde:1995hck}. 
The integrals  in \rf{b4},\rf{b6}   converge if $c>0$. 
For $c <0$   we   define \rf{b7} by an analytic continuation.

\section{Tree-level 6-point amplitude}
\label{a3}

From the expression for  $\mathcal L_6$ in \rf{27},\rf{26}  (parametrized  for generality  by constants $c_4,c_5,c_6$)   we obtain the following three Feynman diagrams contributing to the tree-level amplitude of  scattering of 6 scalars 
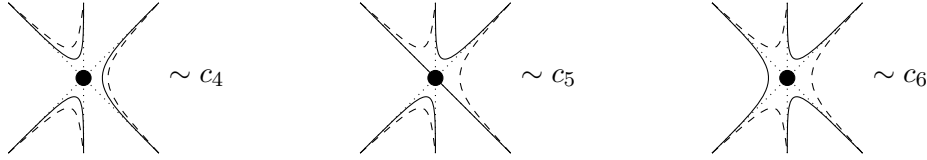
\begin{center}
\begin{tikzpicture}
\draw[dotted] (-1,-1)  -- (0,0);
\draw[dotted] (0,-1) -- (0,0);
\draw[dotted] (1,-1)-- (0,0);

\draw[dotted] (-1,1) -- (0,0);
\draw[dotted] (0,1)  -- (0,0);
\draw[dotted] (1,1)  -- (0,0);

\draw[-] (-1,-1) .. controls (0,0) .. (0,-1); 
\draw[dashed] (-1,-1) .. controls (-0.1,-0.2) .. (0,-1); 

\draw[-] (1,-1) .. controls (0,0) .. (1,1); 
\draw[dashed] (1,-1) .. controls (0.1,0) .. (1,1); 

\draw[-] (-1,1) .. controls (0,0) .. (0,1); 
\draw[dashed] (-1,1) .. controls (-0.1,0.2) .. (0,1); 

\draw[fill] (0,0) circle (0.1);

\draw[] (1,0) node[right] {$\sim c_4$};
\end{tikzpicture} \qquad \qquad
\begin{tikzpicture}
\draw[dotted] (-1,-1)  -- (0,0);
\draw[dotted] (0,-1) -- (0,0);
\draw[dotted] (1,-1)-- (0,0);

\draw[dotted] (-1,1) -- (0,0);
\draw[dotted] (0,1)  -- (0,0);
\draw[dotted] (1,1)  -- (0,0);

\draw[-] (-1,-1) .. controls (0,0) .. (0,-1); 
\draw[dashed] (-1,-1) .. controls (-0.1,-0.2) .. (0,-1); 

\draw[-] (1,-1) .. controls (0,0) .. (-1,1); 
\draw[dashed] (1,-1) .. controls (0.1,0) .. (1,1); 

\draw[-] (1,1) .. controls (0,0) .. (0,1); 
\draw[dashed] (-1,1) .. controls (-0.1,0.2) .. (0,1); 

\draw[fill] (0,0) circle (0.1);
\draw[] (1,0) node[right] {$\sim c_5$};
\end{tikzpicture} \qquad \qquad
\begin{tikzpicture}
\draw[dotted] (-1,-1)  -- (0,0);
\draw[dotted] (0,-1) -- (0,0);
\draw[dotted] (1,-1)-- (0,0);

\draw[dotted] (-1,1) -- (0,0);
\draw[dotted] (0,1)  -- (0,0);
\draw[dotted] (1,1)  -- (0,0);

\draw[-] (-1,-1) .. controls (0,0) .. (-1,1); 
\draw[dashed] (-1,-1) .. controls (-0.1,-0.2) .. (0,-1); 

\draw[-] (0,-1) .. controls (0,0) .. (1,-1); 
\draw[dashed] (1,-1) .. controls (0.1,0) .. (1,1); 

\draw[-] (1,1) .. controls (0,0) .. (0,1); 
\draw[dashed] (-1,1) .. controls (-0.1,0.2) .. (0,1); 

\draw[fill] (0,0) circle (0.1);
\draw[] (1,0) node[right] {$\sim c_6$};
\end{tikzpicture}
\end{center}
Assuming that incoming momenta are  $p_1,p_2,p_3$ and outgoing  ones are $p_1',p_2',p_3'$  the amplitude with the $SO(\hat{D})$ indices  contracted as indicated in  the picture  below 
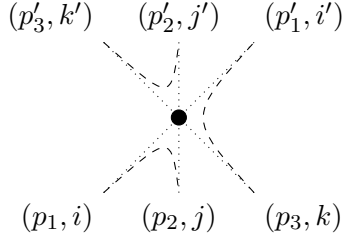
\begin{center}
\begin{tikzpicture}
\draw[dotted] (-1,-1)node[below left] {$(p_1,i)$}  -- (0,0);
\draw[dotted] (0,-1) node[below] {$(p_2,j)$} -- (0,0);
\draw[dotted] (1,-1) node[below right] {$(p_3,k)$}-- (0,0);

\draw[dotted] (-1,1) node[above left] {$(p_3',k')$} -- (0,0);
\draw[dotted] (0,1) node[above] {$(p_2',j')$} -- (0,0);
\draw[dotted] (1,1) node[above right] {$(p_1',i')$} -- (0,0);
 
\draw[dashed] (-1,-1) .. controls (-0.1,-0.2) .. (0,-1); 
 
\draw[dashed] (1,-1) .. controls (0.1,0) .. (1,1); 

\draw[dashed] (-1,1) .. controls (-0.1,0.2) .. (0,1); 

\draw[fill] (0,0) circle (0.1);
\end{tikzpicture} 
\end{center}
contributes a term
\begin{equation} \label{eq:Mtree6}
\mathcal M_{ijk,i'j'k'}\ \  \supset \ \ A\zz[p_1,p_2,p_3,p_1',p_2',p_3'] \, \delta_{ij} \delta_{k i'} \delta_{j'k'} \ , 
%\, \delta^{(d)}(p_1+p_2+p_3-p_1'-p_2'-p_3'),
\end{equation}
\begin{align}
&A\zz[p_1,p_2,p_3,p_1',p_2',p_3'] = 8c_4 (p_1 \cdot p_2) (p_3 \cdot p_1') (p_2' \cdot p_3') + \frac{4}{3} c_5 (p_1 \cdot p_2) (p_3 \cdot p_3')(p_1' \cdot p_2')\no  \\
&\qquad + \frac{4}{3} c_5 (p_1 \cdot p_2) (p_3 \cdot p_2') (p_1' \cdot p_3') 
+ \frac{4}{3} c_5 (p_3 \cdot p_1') (p_1 \cdot p_2')(p_2 \cdot p_3') + \frac{4}{3} c_5 (p_3 \cdot p_1')(p_1 \cdot p_3') (p_2 \cdot p_2')\no  \\
&\qquad+ \frac{4}{3} c_5 (p_2' \cdot p_3')(p_1 \cdot p_3)(p_2 \cdot p_1')+ \frac{4}{3} c_5 (p_2' \cdot p_3')(p_1 \cdot p_1')(p_2 \cdot p_3) 
+ c_6 (p_1 \cdot p_3) (p_2 \cdot p_2') (p_1' \cdot p_3')\no \\  &\qquad  + c_6 (p_1 \cdot p_3) (p_2 \cdot p_3') (p_1' \cdot p_2')
+c_6 (p_1 \cdot p_1') (p_2 \cdot p_2') (p_3 \cdot p_3') + c_6 (p_1 \cdot p_1') (p_2 \cdot p_3') (p_3 \cdot p_2')\no  \\
&\qquad+c_6 (p_1 \cdot p_2') (p_2 \cdot p_3) (p_1' \cdot p_3') + c_6 (p_1 \cdot p_2') (p_2 \cdot p_1') (p_3 \cdot p_3') \no \\
&\qquad +c_6 (p_1 \cdot p_3') (p_2 \cdot p_3) (p_1' \cdot p_2') + c_6 (p_1 \cdot p_3') (p_2 \cdot p_1') (p_3 \cdot p_2'). \label{eq:Atree6}
\end{align}
To get  contributions to $\mathcal M$  with other contractions of indices 
we  need to permute the momenta.

 In  the  case of the effective 2d action of  the compactified membrane \rf{32},
  assuming the ingoing particles  have mode numbers  or  masses
  $(n_1,n_2,n_3)$ and the outgoing ones    $(n_1',n_2',n_3')$,
   the corresponding  6-point amplitude can be obtained from, e.g.,~\eqref{eq:Atree6}   by  the replacement 
\begin{equation}
p_j \cdot p_k \rightarrow p_j \cdot p_k + n_j n_k~.
\end{equation}

\iffa 
\paragraph{6-point from two 4-point.}
We can also construct 6-point processes by combining two 4-point vertices, 
\begin{center}
\begin{tikzpicture}
\draw[dotted] (-2,2)node[left] {$p_{\sigma(1)}$}  -- (-1,0);
\draw[dotted] (-2,0) node[left] {$p_{\sigma(2)}$} -- (-1,0);
\draw[dotted] (-2,-2) node[left] {$p_{\sigma(3)}$}-- (-1,0);

\draw[dotted] (2,2) node[right] {$p_{\sigma(3')}$} -- (1,0);
\draw[dotted] (2,0) node[right] {$p_{\sigma(2')}$} -- (1,0);
\draw[dotted] (2,-2) node[right] {$p_{\sigma(1')}$} -- (1,0);
 
%\draw[dashed] (-2,2) .. controls (-1-0.1,0) .. (-2,0); 
 
%\draw[dashed] (-2,-2) .. controls (-1,0.5) and (1,0.5) .. (2,-2); 

%\draw[dashed] (2,2) .. controls (1+0.1,0) .. (2,0); 

\draw[fill] (-1,0) circle (0.1);
\draw[fill] (1,0) circle (0.1);
\draw[dotted] (-1,0) -- (1,0);

\end{tikzpicture} 
\end{center}
For notation we take that a particle with momentum $p_j$ has target space index $i_j$. In the massive case a mode number $n_j$ also needs to be considered.
There are ten different channels that need to be considered. These can be obtained from the above master geometry through the ten different permutations of the external momenta given by
\begin{equation} \begin{aligned}
\sigma &= (\underline{{\color{blue}1},{\color{blue}2}},\overline{{\color{red}3},{\color{red}1'}},\underline{2',3'}) &\qquad \sigma &= (\underline{{\color{blue}1},{\color{blue}2}},\overline{{\color{red}1'},{\color{red}3}},\underline{2',3'})  & &\\
 \sigma &=(\underline{{\color{blue}1},{\color{blue}2}},\overline{2',3'},\underline{{\color{red}3},{\color{red}1'}}) &\qquad \sigma &= (\underline{{\color{blue}1},{\color{blue}2}},\overline{3',2'},\underline{{\color{red}3},{\color{red}1'}}) \\
\sigma &= (\underline{{\color{red}3},{\color{red}1'}},\overline{{\color{blue}1},{\color{blue}2}},\underline{2',3'}) &\qquad \sigma &=(\underline{{\color{red}3},{\color{red}1'}},\overline{{\color{blue}2},{\color{blue}1}},\underline{2',3'})  \\
\sigma &= (\underline{{\color{red}3},{\color{red}1'}},\overline{2',3'},\underline{{\color{blue}1},{\color{blue}2}}) &\qquad \sigma &= (\underline{{\color{red}3},{\color{red}1'}},\overline{3',2'},\underline{{\color{blue}1},{\color{blue}2}})  \\
\sigma &=(\underline{2',3'},\overline{{\color{blue}1},{\color{blue}2}},\underline{{\color{red}3},{\color{red}1'}})&\qquad \sigma &=(\underline{2',3'},\overline{{\color{blue}2},{\color{blue}1}},\underline{{\color{red}3},{\color{red}1'}})
\end{aligned}
\end{equation}
If two entries are underlined this means that the corresponding target space indices are contracted. If two entries are overlined this means that their target space indices are contracted with the one of the internal leg. We are only interested in those processes where the target space indices are contracted for the particles ${\color{blue}(1,2)}$, ${\color{red}(3,1')}$ and $(2',3')$, as these are the ones that will give rise to a process with the same index structure as the highlighted term in \eqref{eq:Mtree6}. Assuming that all particles are incoming, from momentum conservation we deduce that the momentum in the internal leg is given by
\begin{equation}
p = p_{\sigma(1)}+p_{\sigma(2)}+p_{\sigma(3)}.
\end{equation}
We then need to compute the Feynman diagram, assuming again all momenta are incoming
\begin{equation} 
\sum_\sigma \left. \frac{A^{(0)}[p_{\sigma(1)}, p_{\sigma(2)}, p_{\sigma(3)}, p]A^{(0)}[p,p_{\sigma(1')}, p_{\sigma(2')}, p_{\sigma(3')}]}{p^2 - i \varepsilon} \right|_{p=p_{\sigma(1)}+p_{\sigma(2)}+p_{\sigma(3)}}
\end{equation}
We then need to put all the contributions on the same denominator.

Notice that we can describe any type of process $\mathcal M_{n \rightarrow m}$ with $n+m=6$. To change some particles from ingoing into outgoing we just need to change the sign of the energy-momentum.
\fi 

\section{Details of 1-loop    computation in section \ref{s33} }
%Fixed $n$ calculation}
\label{a4}

The   computation  of  the 1-loop  amplitude in the case of the compactified   membrane 
in section \ref{s33} 
follows the discussion in the NG case (see \rf{221},\rf{223},\rf{2255}). 

 For example, in the $s$-channel  we need to compute the integral 
\begin{equation}\la{d1} 
\mathcal I_s \equiv  \int_0^1 dx\,  I_s =  \int_0^1 d x \, \Big( \GG_{0,s} \Gamma(1+\epsilon) \,  \dDelta_s^{-1-\epsilon}
 + \GG_{2,s} \Gamma(\epsilon) \, \dDelta_s^{-\epsilon}+ \GG_{4,s} \Gamma(-1+\epsilon)\, \dDelta_s^{1-\epsilon}\Big)\ , 
\end{equation} 
where  $\gamma_{r,s}$  are given in \eqref{3223} and $\dDelta_s = n^2 - x (1-x) s$
as in \rf{318}. %The expression for the other channels is similar.  
 In the limit  $\epsilon \rightarrow 0$ we get 
\begin{equation}\la{d2}
\mathcal I_{s} = \int_0^1 d x \Big[ \GG_{0,s} \, {\dDelta^{-1}_s} + \GG_{2,s} \Big( \frac{1}{\epsilon} - \gamma + \ln 4 \pi - \ln \dDelta_s\Big) + \GG_{4,s} \Big(\frac{1}{\epsilon} - \gamma + \ln 4 \pi - \ln \dDelta_s +1  \Big) (-\dDelta_s ) \Big].
\end{equation}
Adding  similar contributions of the  other   channels  using \rf{32231},\rf{32232}, the total integrand  may be written as 
\begin{equation} \begin{aligned}\la{d3}
I &=I_s+I_t+I_u=\frac{1}{4}(\hat D -8)    s^4 x^2 (1-x)^2 \, {\dDelta^{-1}_s}
 + 2 t^4 x^2 (1-x)^2 \frac{1}{\dDelta_{t}} + 2 u^4 x^2 (1-x)^2 \, {\dDelta^{-1}_{u}}  \\
&+ \big[-s^3 x (1-x)\big]  \Big( \frac{1}{\epsilon} - \gamma + \ln 4 \pi - \ln \dDelta_s\Big) + \big[s^2 -\frac{1}{2} \hat D\, u t \big] \Big(\frac{1}{\epsilon} - \gamma + \ln 4 \pi - \ln \dDelta_s +1  \Big) (-\dDelta_s )\\
& + \big[\frac{1}{2} t^2 s + 3 t^3 x(1-x) \big]\Big( \frac{1}{\epsilon} - \gamma + \ln 4 \pi - \ln \dDelta_{t}\Big)+ \big[\frac{1}{2} u^2 s + 3 u^3 x(1-x) \big]\Big( \frac{1}{\epsilon} - \gamma + \ln 4 \pi - \ln \dDelta_{u}\Big) . \\
\end{aligned}
\end{equation}
For the divergent part  we find
\begin{equation} \begin{aligned}
I_\epsilon &=\Big( \frac{1}{\epsilon} -\gamma  + \ln 4 \pi\Big) \Big(  \big[ \frac{1}{2} - 3 x(1-x)\big]
 s^3 -\big[1+ \frac{1}{2} (\hat D -18)  x (1-x)\big]   s t u - \big[s^2 - \frac{1}{2}\hat D\,  u t \big] n^2  \Big).
\end{aligned}
\end{equation}
Using that $\int_0^1 x(1-x) = \frac{1}{6}$ this gives
\begin{equation}\la{d5}
\mathcal I_\epsilon = \int_0^1 dx \, I_\epsilon = \Big(\frac{1}{\epsilon} - \gamma + \ln 4 \pi \Big) \Big[ - \frac{1}{12} (\hat{D}-8) s t u - \frac{n^2}{2} \big( 2 s^2 - \hat{D} u t \big) \Big].
\end{equation}
To simplify the expression  for the finite part   we choose the 
  kinematics  so that $t=0$ and $u=-s$. Then 
\begin{equation} \begin{aligned}\la{d6}
\mathcal I_f &= - \frac{1}{24}(\hat{D}-24) 
 s^3 + s^2 n^2 \Big[ - \frac{1}{4}(\hat{D}+4) 
  + \ln n^2 + \big( \frac{1}{4} \hat{D}\, 
  n^2 - \frac{s}{2} \big) Q_n(-s) - \frac{s}{2} Q_n(s) \Big]\ , 
\end{aligned}
\end{equation}
where 
\be 
Q_n(s)\equiv  \int_0^1 d x \, (\dDelta_{-s})^{-1}  = - \frac{2}{s \sqrt{1+\frac{4 n^2}{s}}} \ln   \frac{ \sqrt{1+\frac{4 n^2}{s}}-1}{\sqrt{1+\frac{4 n^2}{s}}+1}\ , \la{d8} \ee
 and we also  used    the  following 
 integrals\foot{Here the integrands involve   $\dDelta_{-s}= n^2 + x (1-x) s$ so that the resulting 
 expressions   are  well defined for $s>0$. To write $Y(s)$ and $Z(s)$ in terms of $W(s)$ and $Q_n (s)$ we used integration by parts.} 
\begin{align}
W(s) &=\int_0^1 dx \, \dDelta_{-s} = n^2 +\frac{s}{6} \la{d7}\ ,  \qquad 
Y(s)=\int_0^1 dx \, \ln \dDelta_{-s} =  \ln n^2 - 2 + \frac{1}{2} (4 n^2+s) Q_n(s)\ , \\
Z(s)&=\int_0^1 dx \, \dDelta_{-s} \ln \dDelta_{-s} = \Big(n^2 + \frac{s}{6} \Big) Y(s) - \frac{2}{3} W(s) + \frac{8 n^2+s}{6} - \frac{n^2}{6}(4 n^2+s) Q_n(s)\ . \la{d10}
\end{align}

\iffa 

%%%%%%%%%%%%%%%%%%%%%%%%%%%%%%%%%%
\section{$T\bar{T}$ deformation of a massive scalar theory ????????}\la{a5}

Let us consider the case when all fields have the same mass $m$ and start with the seed Lagrangian describing free massive fields,
\begin{equation}
\mathcal L_0 = \frac{1}{2} \Big( \partial_\alpha X^j \partial^\alpha X^j + m^2 X^j X^j \Big).
\end{equation}
To lowest order, the $T\bar{T}$ operator produces a quadratic interaction term
\begin{equation}
O_{T \bar{T}} = -\frac{1}{2} \Big( c_2 \partial_\alpha X^j \partial^\alpha X^j \partial_\beta X^k \partial^\beta X^k + c_3 \partial_\alpha X^j \partial^\beta X^j \partial_\beta X^k \partial^\alpha X^k + c_4 m^4 X^j X^j X^k X^k \Big).
\end{equation}
In this section we consider the scattering of four particles with the same mass $m$ with the action given by the $T\bar{T}$ deformed theory $\mathcal L_0 + \lambda O_{T \bar{T}}$. The issue we encounter here is that UV divergences appear when computing the 4 point amplitude using the classical Lagrangian. The divergent part of the amplitude can be cancelled by adding appropriate counterterms to the Lagrangian, but this leads to ambiguous finite contributions. A way forward is to fix the ambiguity by requiring that the theory is integrable also at the quantum level. 

Such a computation has been done in \cite{Rosenhaus:2019utc}, using a hard cutoff. We here reproduce the calculation using dimensional regularisation.

We find a divergent part
\begin{equation} \begin{aligned}
A_\epsilon &=  -\frac{1}{96 \pi} \Big( (\hat{D}-6) s t u + 6 m^2 (2  s^2-  \hat{D}  t u) - 36 m^4 s  +12 \hat{D} m^6 \Big),
\end{aligned}
\end{equation}
from which we can also deduce
\begin{equation}
B_\epsilon = A_\epsilon (s \leftrightarrow t), \qquad C_\epsilon = A_\epsilon(s \leftrightarrow u)
\end{equation}
In the kinematical region where $t=0$, the finite contribution to the amplitudes $A$ and $C$ read
\begin{equation} \begin{aligned}
A_f &= -\frac{\hat{D}-24}{192 \pi} s^3 + \frac{m^2 s^2}{96 \pi} \Big( 7 \hat{D}-108+ 12 \log m^2 \Big) \\
&\qquad+ \frac{m^4 s}{24 \pi} \Big( 60 -8 \hat{D} - 9 \log m^2 \Big) + \frac{m^6}{8 \pi} \Big( -4 + 3 \hat{D} + \hat{D} \log m^2 \Big)
\end{aligned}
\end{equation}
\begin{equation} \begin{aligned}
C_f &= +\frac{\hat{D}-24}{192 \pi} s^3 + \frac{m^2 s^2}{96 \pi} \Big(\hat{D}+ 36 + 12 \log m^2 \Big) \\
&\qquad+ \frac{m^4 s}{24 \pi} \Big( 4 - 5\log m^2 \Big) + \frac{m^6}{8 \pi} \Big( -4 - \hat{D} + (\hat{D}+4) \log m^2 \Big)
\end{aligned}
\end{equation}
In particular, these expressions do not depend on $X(s)$. The finite part of $A$ and $C$ is therefore polynomial in $s$, and can be cancelled by an appropriate counterterm added to the classical Lagrangian. For $m \rightarrow 0$ we recover the Nambu-Goto result with $\hat{D}=D-2$ the number of Goldstone modes. For the amplitude $B$ on the other hand, $X(s)$ does not cancel out, and we are left with
\begin{equation} \begin{aligned}
B_f &= \frac{s^2 (4 m^2-s)^2}{32 \pi} \Big( X(-s) + X(-4 m^2+s)\Big) \\
&\qquad+ \frac{m^2 s^2}{16 \pi} \Big( 4 - \hat{D}+\hat{D} \log m^2  \Big) - \frac{m^4 s}{4 \pi} \Big(4-\hat{D}+\hat{D} \log m^2 \Big) + \frac{m^6}{8 \pi} \Big(-4 + 3 \hat{D} + \hat{D} \log m^2 \Big)
\end{aligned}
\end{equation}
where we recall that
\begin{equation}
X(s) = \frac{2}{s \sqrt{1+\frac{4 m^2}{s}}} \ln \frac{\sqrt{1+\frac{4 m^2}{s}}+1}{\sqrt{1+\frac{4 m^2}{s}}-1}.
\end{equation}
In fact, using the notation $p_1 = m \sinh \theta_1$ and $p_2 = m \sinh \theta_2$ we have that $s = 2 m^2(1+\cosh(\theta_1-\theta_2))$ and
\begin{equation}
X(-s) + X(-4 m^2+s) = -\frac{i \pi}{m^2 \sinh (\theta_1-\theta_2)}
\end{equation}
and therefore
\begin{equation}
B_f = -\frac{i}{2} m^6 \sinh^3(\theta_1-\theta_2) + \dots
\end{equation}
where the dots are real terms, polynomial in $s$.

In the case where $\hat{D}=1$ there is only one scalar and the total amplitude reads
\begin{equation}
A_f+B_f+C_f = 
\end{equation}

\begin{equation} \begin{aligned}
A_\epsilon &= -\frac{1}{12} \Big( \hat{D} \Big( s t u  + 12 m^4 n^2 - 6 n^2 t u  \Big) +6  s \Big( - t u - 6 m^4 + 2 m^2 s \Big) \Big)\\
&= -\frac{1}{12} \Big( (\hat{D}-6) s t u - 6 \hat{D} n^2 t u - 36 m^4 s + 12 m^2 s^2 +12 \hat{D} m^4 n^2 \Big)
\end{aligned}
\end{equation}

\fi

\

\bibliographystyle{JHEP-v2.9}
\small 
\bibliography{biblio}

\end{document}